\documentclass[11pt]{article}
\usepackage{aas_macros,amsmath,amssymb,comment,cite,esint,graphicx,mathtools,diagbox}
\usepackage{bm}
\usepackage[margin=.8in,letterpaper]{geometry}
\usepackage[colorlinks=true]{hyperref}
\usepackage[affil-it]{authblk}
\usepackage{subcaption}
\usepackage[utf8]{inputenc}

\usepackage{mathrsfs}
\usepackage{appendix}
\usepackage{amssymb}
\usepackage{float}                  
\usepackage{color}
\usepackage{cite}
\usepackage{hyperref}
\hypersetup{pageanchor=false}
\usepackage{indentfirst}
\usepackage{url}
\usepackage{xfrac}
\usepackage{caption}
\usepackage[numbers,square,comma,sort&compress,merge]{natbib}
\usepackage{esint}
\usepackage{overpic}
\usepackage{graphicx}
\usepackage{epsf,amsmath,bbold,amsfonts,stmaryrd}
\usepackage{textcomp}
\usepackage{ulem}
\usepackage{tikz}
\usepackage{multirow}
\numberwithin{equation}{section}
\let\originalleft\left
\let\originalright\right
\renewcommand{\left}{\mathopen{}\mathclose\bgroup\originalleft}
\renewcommand{\right}{\aftergroup\egroup\originalright}

\def\bea{\begin{eqnarray}}
\def\eea{\end{eqnarray}}
\def\nn{\nonumber}

\usepackage{tensor}
\usepackage{physics}

\newcolumntype{P}[1]{>{\Centering\hspace{0pt}}p{#1}}
\newcolumntype{Z}{>{\centering\arraybackslash}X} 

\newcommand{\df}{\mathrm{d}}   
   
\begin{document}
\title{\bf Imaging and polarization patterns of various thick disks around Kerr-MOG black holes }
	
\author{Xinyu Wang$^{1,2}$, Huan Ye$^{3}$, Xiao-Xiong Zeng$^{4\ast}$}
\date{}
	
\maketitle
\vspace{-15mm}

\begin{center}
{\it
$^1$ School of Physics and Astronomy, Beijing Normal University,
Beijing 100875, P. R. China\\\vspace{2mm}

$^2$Key Laboratory of Multiscale Spin Physics (Ministry of Education), Beijing Normal University, Beijing 100875,  P. R. China\\\vspace{2mm}

$^3$ School of Material Science and Engineering, Chongqing Jiaotong University, Chongqing 400074, P. R. China\\\vspace{2mm}

$^4$ College of Physics and Electronic Engineering,
Chongqing Normal University, Chongqing 401331,  P. R. China \\\vspace{2mm}
}
\end{center}

\vspace{8mm}

\begin{abstract}
We investigate the imaging and polarization properties of Kerr–MOG black holes surrounded by geometrically thick accretion flows. The MOG parameter $\alpha$ introduces deviations from the Kerr metric, providing a means to test modified gravity in the strong field regime. Two representative accretion models are considered: the phenomenological radiatively inefficient accretion flow (RIAF) and the analytical ballistic approximation accretion flow (BAAF). Using general relativistic radiative transfer, we compute synchrotron emission and polarization maps under different spins, MOG parameters, inclinations, and observing frequencies. In both models, the photon ring and central dark region expand with increasing $\alpha$, whereas frame dragging produces pronounced brightness asymmetry. The BAAF model predicts a narrower bright ring and distinct polarization morphology near the event horizon. By introducing the net polarization angle $\chi_{\text{net}}$ and the second Fourier mode $\angle\beta_2$, we quantify inclination- and frame-dragging–induced polarization features. Our results reveal that both $\alpha$ and spin significantly influence the near-horizon polarization patterns, suggesting that high-resolution polarimetric imaging could serve as a promising probe of modified gravity in the strong field regime.

\end{abstract}

\vfill{\footnotesize $\ast$ Corresponding author: xxzengphysics@163.com}

\maketitle

\newpage
\baselineskip 18pt

\section{Introduction}\label{sec1}

The existence of black holes is now firmly established through multiple, independent observations, including the detection of gravitational waves by the LIGO/Virgo Collaborations~\cite{LIGOScientific:2016aoc} and the horizon-scale images of M87* and Sgr~A* captured by the Event Horizon Telescope (EHT)~\cite{EventHorizonTelescope:2019dse,EventHorizonTelescope:2022wkp}. More recently, EHT polarization measurements~\cite{EventHorizonTelescope:2021bee,EventHorizonTelescope:2024hpu} have provided unprecedented insights into the magnetized plasma and radiation processes operating near the event horizon, offering powerful probes of both accretion physics and spacetime geometry. These advances have stimulated widespread interest in the imaging of black hole accretion disks, particularly in polarized emission, within both general relativity (GR)~\cite{Hou:2023bep,Zhang:2024lsf,Zhang:2025vyx} and various modified gravity frameworks~\cite{Yang:2025usj,Zeng:2025pch,Wang:2025edt,Zeng:2025kyv,Yang:2025whw,Wang:2025buh,Chen:2025ysv,Li:2025knj,Hou:2022eev}.

Although general relativity (GR) has been extensively validated~\cite{Will:2014kxa}, deviations at galactic and cosmological scales remain possible. Observed discrepancies in galaxy rotation curves~\cite{1965ApJ...141..885R,1970ApJ...159..379R} and cluster dynamics are often attributed to dark matter, whose direct detection remains elusive. An alternative approach is to modify gravity itself. One such theory is the Scalar–Tensor–Vector Gravity (STVG), or Modified Gravity (MOG), proposed by Moffat~\cite{Moffat:2005si}. In this framework, gravity is mediated by three fields: the Einstein metric corresponding to a massless tensor graviton, a massless scalar field, and a massive Proca-type vector field. The vector field coupled to a scalar field that dynamically determines the effective gravitational coupling strength, $G = G_N(1 + \alpha)$, where $\alpha$ is a dimensionless MOG parameter. This enhancement strengthens gravity at galactic and cosmological scales, successfully reproducing galaxy rotation curves~\cite{Moffat:2013sja,Moffat:2014pia,Moffat:2023ffv} and cluster dynamics~\cite{Brownstein:2007sr,Moffat:2006ii,Moffat:2013uaa} without invoking dark matter.

The STVG/MOG framework also admits stationary, axisymmetric black-hole solutions that generalize the Kerr metric via an $\alpha$-dependent modification of the gravitational potential~\cite{Moffat:2014aja}. The MOG parameter introduces a new degree of freedom affecting various black-hole properties, such as thermodynamics~\cite{Mureika:2015sda,Cai:2020igv}, particle dynamics near the horizon~\cite{Sharif:2017owq,Lee:2017fbq,Sheoran:2017dwb,Haydarov:2020xnv,Rayimbaev:2020lyg,Murodov:2023one}, and quasinormal modes~\cite{Manfredi:2017xcv,Liu:2023uft}. Gravitational waves within STVG/MOG have also been explored~\cite{Moffat:2016gkd,Green:2017qcv}. More recently, analytic studies of force-free magnetic fields and Blandford–Znajek jet emission in Kerr–MOG background have been carried out~\cite{Camilloni:2023wyn}. Moreover, observational features of Kerr–MOG black holes—including shadow structures~\cite{Moffat:2015kva,Wang:2018prk,Hu:2022lek} and accretion-disk emission—have attracted growing attention. In particular, within the thin-disk model, the parameter $\alpha$ has been found to influence both the image brightness and morphology~\cite{Perez:2017spz,Hu:2023bzy}, as well as the polarization properties of emitted radiation~\cite{Qin:2022kaf}.

Supermassive black holes (SMBHs) are generally understood to accrete hot, magnetized plasma, giving rise to luminous accretion disks \cite{Lu:2023bbn,EventHorizonTelescope:2019pgp}. Owing to the strong gravitational field, the suppression of vertical cooling together with the compression of inflowing material~\cite{Narayan:1994xi} tends to produce disks that are geometrically thick and optically thin~\cite{EventHorizonTelescope:2019dse,EventHorizonTelescope:2019pgp,EventHorizonTelescope:2022urf,Ho:1999ss}. In such environments, a detailed characterization of the streamlines, particle number density, electron temperature, and magnetic-field structure is required. The physics of geometrically thick accretion disks therefore remains an active and ongoing area of theoretical research~\cite{Abramowicz:2011xu}. 

A widely used phenomenological framework for modeling geometrically thick accretion flows is the \textit{radiatively inefficient accretion flow} (RIAF) model~\cite{Yuan:2003dc,Broderick:2005at,Broderick:2008qf,Broderick:2010kx,Pu:2016qak,Pu:2018ute,Jiang:2023img,Saurabh:2025kwb,Yin:2025rao}, 
in which the vertically averaged electron density and temperature exhibit an approximate power-law dependence on the radius~\cite{Yuan:2003dc}. Another representative approach is the torus model introduced by Abramowicz \textit{et al.}~\cite{1978A&A....63..221A}, which was subsequently extended to include magnetized tori~\cite{Komissarov:2006nz}. The Kerr-MOG black hole images illuminate by such geometrically thick magnetized equilibrium tori and to probe effects of the MOG parameter on the images has been studied in \cite{Zhang:2024jrw}. These disk models predominantly focus on nearly circular orbits and large-scale behaviors. Motivated by the possibility of horizon-resolved black hole observations, Hou \textit{et al.}~\cite{Hou:2023bep,Zhang:2024lsf} recently developed an analytical \textit{ballistic approximation accretion flow} (BAAF) model within the general  relativistic magnetohydrodynamics (GRMHD) framework. In this approach, magnetofluids dominated by gravity follow ballistic geodesics, allowing explicit thermodynamic and magnetic-field profiles to be obtained, which describe both thick disks and jets at horizon scales. This framework provides a valuable tool for black hole imaging and the study of near-horizon polarization.

In this work, we investigate the imaging properties of Kerr–MOG black holes using two representative accretion models: the phenomenological RIAF and the analytical BAAF. We compute images using a general relativistic radiative transfer (GRRT) approach and analyze the resulting intensity morphology under varying parameters, inclinations, and observing frequencies, including the impact of anisotropic synchrotron emission. Within the BAAF framework, we further explore the polarization signatures, introducing the net polarization $\chi_\text{net}$ to quantify inclinations dependence and $\angle\beta_2$ to characterize frame-dragging effects and intrinsic polarization vectors near the event horizon.

The remaining sections of this paper are structured as follows.
In Sec.~\ref{sec2}, we present the essential
 concepts of black hole spacetimes, the flow dynamics considered in this work, and the synchrotron radiation mechanism and the GRRT methodology.
In Sec.~\ref{sec3}, we analyze the imaging characteristics under the RIAF model, including the effects of emission anisotropy and observation frequency.
Sec.~\ref{sec4} focuses on the BAAF model and presents the resulting intensity and polarization patterns.
Finally, Sec.~\ref{sec5} presents a summary and outlook. In this work, we have set the fundamental constants $c$, $G$ to unity, unless otherwise specified, and we will work in the convention $(-,+,+,+)$.

\section{Preliminary and method}\label{sec2}
In this section, we present the essential concepts related to the Kerr–MOG spacetime, timelike and null geodesics, and the methodologies employed for imaging and polarization studies. The specific accretion flow models will be discussed in the subsequent sections.

\subsection{Kerr-MOG black hole and geodesics}
\label{metric}

The STVG theory~\cite{Moffat:2005si} is a covariant modified theory of gravity incorporating scalar, tensor, and vector fields. Its total action can be expressed as 
\begin{equation}
S = S_{\rm GR} + S_\phi + S_S + S_M\,,
\end{equation}
where
\begin{align}
S_{\rm GR} &= \frac{1}{16\pi} \int d^4x \sqrt{-g} \frac{R}{G}\,, \\
S_\phi &= \int d^4x \sqrt{-g} \left( -\frac{1}{4} B^{\mu\nu}B_{\mu\nu} + \frac{1}{2}\mu^2 \phi^\mu \phi_\mu \right)\,, \\
S_S &= \int d^4x \sqrt{-g} \left[ \frac{1}{G^3} \left( \frac{1}{2} g^{\mu\nu} \nabla_\mu G \nabla_\nu G - V(G) \right) + \frac{1}{\mu^2 G} \left( \frac{1}{2} g^{\mu\nu} \nabla_\mu \mu \nabla_\nu \mu - V(\mu) \right) \right]\,.
\end{align}
Here, $S_S$ corresponds to the GR action and $R$ is the Ricci scalar, $S_\phi$ is the action of the Proca-type massive vector field $\phi^\mu$ and $B_{\mu\nu}=\partial_\mu \phi_\nu-\partial_\nu \phi_\mu$, and $G$ is a scalar field corresponding to a spin $0$ massless graviton. In this modified theory, the mass of the vector field $\phi^\mu$ is also an effective spin $0$ scalar field $\mu(x)$. $V(G)$ and $V(\mu)$ are self-interaction potentials of the $G(x)$ and $\mu(x)$ fields, respectively. $S_M$ denotes the matter action. The mass scale $\mu(x)$ controls the effective range of the vector field, which becomes relevant on kiloparsec scales. Since long-range effects are negligible in black hole solutions, the vector mass $\mu$ can be safely omitted. In vacuum~\cite{Wang:2025zis}, $G$ can be treated as a constant, independent of the spacetime coordinates. Under these assumptions, the action then simplifies to
\begin{equation}
S = \int d^4x \sqrt{-g} \left( \frac{R}{16\pi G} - \frac{1}{4} B^{\mu\nu}B_{\mu\nu} \right).
\end{equation}

The effective gravitational constant is related to the Newtonian one via $G = G_N(1 + \alpha)$, where $\alpha$ is a dimensionless parameter quantifying deviations from GR. This enhancement of $G$ reflects the scale-dependent nature of gravity in MOG, offering a mechanism to explain galactic rotation curves, lensing effects, and large-scale structure without invoking dark matter. When $\alpha = 0$, the theory reduces exactly to GR, making $\alpha$ a key parameter for theoretical modeling and observational tests. We refer to $\alpha$ as the MOG parameter throughout. Applying the modified Newman–Janis algorithm, the spacetime of a rotating black hole in the STVG theory can be described by the so-called Kerr–MOG metric with the form in Boyer–Lindquist (BL) coordinate \cite{Moffat:2014aja}:
\begin{equation}
\begin{aligned}
ds^2 = & -\frac{\Delta - a^2 \sin^2\theta}{\Sigma} dt^2 - 2a \sin^2\theta \left( \frac{r^2 + a^2 - \Delta}{\Sigma} \right) dt d\phi \\
& + \sin^2\theta \left( \frac{(r^2 + a^2)^2 - \Delta a^2 \sin^2\theta}{\Sigma} \right) d\phi^2 + \frac{\Sigma}{\Delta} dr^2 + \Sigma d\theta^2\,,
\end{aligned}
\end{equation}
where
\begin{align}
\Delta &= r^2 - 2GMr + a^2 + \alpha G_N G M^2\,,\quad
\Sigma = r^2 + a^2 \cos^2\theta\,.
\end{align}
Here $M$ and $a$ denote the black hole mass and spin, respectively. Furthermore, for the sake of simplicity and without loss of generality, we shall set $M = G_N=1$ in the forthcoming discussion. The Proca-type vector field introduces the $\alpha$-dependent term in $\Delta$, thereby modifying the gravitational potential. 

The outer and inner horizon radii are the roots of $\Delta=0$,
\begin{equation}
r_\pm = 1+\alpha\pm \sqrt{1+\alpha-a^2}\,,
\end{equation}
with the outer horizon radius increasing monotonically with $\alpha$. The extremal limit occurs for
\begin{equation}
    \alpha=a^2-1\,.
\end{equation}

The geodesic equations are completely integrable since they admit four constants along the trajectory of a particle: \(g_{\mu\nu}u^\mu u^\nu = -m^2\), with \(m^2 = 0\) and \(m^2 = 1\) for null and timelike geodesics, respectively; the energy \(E = -u \cdot \partial_t\); the angular momentum \(L = u \cdot \partial_\phi\); and the Carter constant \(Q\). Here, we have assumed \(u^\mu = \frac{dx^\mu}{d\tau}\), where \(u^\mu\) is interpreted as the four-velocity or the four-momentum per unit mass for timelike geodesics, with \(\tau\) representing the proper time. The physical meanings of \(E\) and \(L\) in this context correspond to the energy and angular momentum per unit mass, respectively. For null geodesics, \(u^\mu\) denotes the photon's four-wavevector, and \(\tau\) should be understood as the affine parameter along the worldline. In this case, \(E\) and \(L\) are independent of mass. By utilizing these conserved quantities, the geodesic equations can be written in first-order form \cite{Bardeen:1973tla}:
\bea  
\label{fv}  
u^r & = & \pm_r \frac{1}{\Sigma} \sqrt{\mathcal{R}(r)} \,, \nn\\  
u^\theta & = & \pm_\theta \frac{1}{\Sigma} \sqrt{\Theta(\theta)} \,, \nn\\  
u^\phi & = & \frac{1}{\Sigma} \left[\frac{a}{\Delta} \left(E\left(r^2 + a^2\right) - a L\right) + \frac{L}{\sin^2 \theta} - a E\right] \,, \nn\\  
u^t & = & \frac{1}{\Sigma} \left[\frac{r^2 + a^2}{\Delta} \left(E\left(r^2 + a^2\right) - a L\right) + a\left(L - a E \sin^2 \theta\right)\right] \,.  
\eea  
Here,  
\bea\label{pots}  
\mathcal{R}(r) & = & \left[E\left(r^2 + a^2\right) - a L\right]^2 - \Delta \left[Q + (L - a E)^2 + m^2 r^2\right] \,, \nn\\  
\Theta(\theta) & = & Q + a^2\left(E^2 - m^2\right) \cos^2 \theta - L^2 \cot^2 \theta \,,  
\eea  
are the radial and angular potentials, respectively. \(\pm_r\) and \(\pm_\theta\) denote the signs of the radial and polar motions. From the radial potential $\mathcal{R}\geq0$ in Eq. \eqref{pots}, it is evident that timelike geodesic
particles at infinity satisfy $E\geq1$.

In this work, we restrict attention to fluid motion satisfying $u^\theta=0$. In this case, the streamline preserves a constant value of $\theta$ along the polar direction. This requires $\Theta(\theta)=\partial_\theta \Theta(\theta)=0$, leading to 
\begin{equation}
\label{signl}
L= \pm_L a \sqrt{E^2-1} \sin ^2 \theta, \quad Q=-a^2\left(E^2-1\right) \cos ^4 \theta\,,
\end{equation}
where ``$\pm_L$'' denotes the sign of $L$. The negative sign of $L$ corresponds to an accretion flow rotating in the direction opposite to the black hole spin, 
while the positive sign indicates co-rotation with the black hole. Furthermore, the condition
\begin{equation}
\partial_{\theta}^{2}\Theta = -8a^{2}(E^{2}-1)\cos^{2}\theta \leq 0
\end{equation}
must be satisfied to ensure the stability of the geodesics. 
Since the fluid is foliated by streamlines lying on conical surfaces, 
we refer to the solution satisfying Eq.\eqref{signl} as the conical solution. The radial potential can be expressed as
\begin{equation}
\begin{aligned}
    \mathcal{R}_\text{c}(r,\theta)&=(E^2-1)r^4+2(1+\alpha)r^3+\left[2a^2(E^2-1)\cos^2{\theta}-\alpha(1+\alpha)\right]r^2\\
    &+2a^2(1+\alpha)\left[(E \mp_L \sqrt{E^2-1} \sin ^2 \theta)^2-\left(E^2-1\right) \cos ^4 \theta\right]r\\
    &+a^2\left[-\alpha(1+\alpha)(E\mp_L \sqrt{E^2-1}\sin^2{\theta})^2-(a^2+\alpha+\alpha^2)(E^2-1)\cos^4{\theta}\right]\,.
\end{aligned}
\end{equation}
Thus, the $4$-velocity takes the form
\begin{equation}
\label{conical}
    \begin{aligned}
        u^r&=\pm_r \frac{1}{\Sigma}\sqrt{\mathcal{R}_\text{c}}\,,\quad u^\theta=0\,,\\
        u^\phi &=E\frac{a(a^2+r^2-\Delta)}{\Delta\Sigma}\pm_L \sqrt{E^2-1}\frac{a(\Delta-a^2\sin^2{\theta})}{\Delta\Sigma}\,,\\
        u^t &=E\frac{(a^2+r^2)^2-a^2\Delta \sin^2{\theta}}{\Delta\Sigma}\mp_L \sqrt{E^2-1}\frac{a^2(a^2+r^2-\Delta)\sin^2{\theta}}{\Delta\Sigma}\,.
    \end{aligned}
\end{equation}
When $L<0$, although accretion flow rotates in an opposite direction
of BH spin, it will be dragged back by gravity. The turning radius, where the $u^\phi$ flip its sign, is:
\begin{equation}
\label{rf}
    r_f=1+\alpha+\frac{(1+\alpha)E}{\sqrt{E^2-1}}+\sqrt{\frac{(1+\alpha)\left[-1+E(E+\sqrt{E^2-1})(2+\alpha)\right]}{E^2-1}-a^2 \cos^2{\theta}}\,.
\end{equation}
As $E$ increases, $r_f$ decreases and approaches an asymptotic value for $E \to \infty$,
\begin{equation}
    r_f=2+2\alpha+\sqrt{4+6\alpha+2\alpha^2}\,.
\end{equation}
evaluated here on the equatorial plane, $\theta=\pi/2$.

A special subclass of the conical solution corresponds to infalling motion with $E=1$, representing fluid is at rest at spatial infinity:
\begin{equation}
\label{infalling}
    u^\mu=(-g^{tt},-\sqrt{-(1+g^{tt})g^{rr}},0,-g^{t\phi})\,.
\end{equation}

\subsection{Synchrotron Radiation and Radiative Transfer}

In this work, we consider the presence of magnetic fields near the black hole, where highly relativistic electrons in the plasma emit synchrotron radiation under the influence of the Lorentz force. All physical quantities are expressed in CGS units.

To facilitate the description of the emission, absorption, and propagation of polarized radiation in curved spacetime, we adopt a suitable orthonormal reference frame~\cite{Broderick:2003fc}.
The corresponding orthonormal basis vectors are given by:
\begin{equation}
e_{(0)}^\mu=u^\mu, \quad e_{(3)}^\mu=\frac{k^\mu}{\omega}-u^\mu, \quad e_{(2)}^\mu=\frac{1}{\mathcal{F}}\left(b^\mu+\beta u^\mu-C e_{(3)}^\mu\right), \quad e_{(1)}^\mu=\frac{\epsilon^{\mu \nu \sigma \rho} u_\nu k_\sigma b_\rho}{\omega \mathcal{F}}\,,
\end{equation}
where
\begin{equation}
b^2=b_\mu b^\mu, \quad \beta=u_\mu b^\mu, \quad \omega=-k_\mu u^\mu, \quad C=\frac{k_\mu b^\mu}{\omega}-\beta, \quad \mathcal{F}=\sqrt{b^2+\beta^2-C^2}\,.
\end{equation}
Here, $\epsilon^{\mu \nu \sigma \rho}$ is the Levi-Civita tensor. In this frame, the emission, absorption, and Faraday rotation coefficients associated with the Stokes parameter $U$ vanish identically. The remaining nonzero coefficients, corresponding to the Stokes parameters $I$, $Q$, and $V$, are approximated as~\cite{Huang:2024bar,Dexter:2016cdk,1980gbs..bookQ....M}:
\begin{equation}
    \begin{aligned}
        j_I&=\frac{n_e e^2\nu}{2\sqrt{3}c\Theta_e^2}I_I(x)\,,\\
        j_Q&=\frac{n_e e^2\nu}{2\sqrt{3}c\Theta_e^2}I_Q(x)\,,\\
        j_V&=\frac{2n_e e^2\nu\cot{\theta_B}}{3\sqrt{3}c\Theta_e^3}I_V(x)\,,
    \end{aligned}
\end{equation}
with 
\begin{equation}
    x\equiv \frac{\nu}{\nu_c}\,,\quad \nu_c = \frac{3 e B \sin{\theta_B} \Theta_e^2}{4\pi m_e c}\,,\quad \Theta_e = \frac{k_B T_e}{m_e c^2}\,.
\end{equation}
where $\nu$ is the frequency of the emitted photons, and $\nu_c$ is the characteristic frequency of the system. $n_e$ is the electron number density and $\Theta_e$ is the dimensionless electron temperature. Here and in what follows, $k_B$ is the Boltzmann constant and $T_e$ the thermodynamic temperature. $B$ represents the local magnetic magnitude. The pitch angle $\theta_B$ is the angle between the wave vector and the magnetic field in the fluid rest frame. $e$, $m_e$, and $c$ are the elementary charge, electron mass, and speed of light, respectively. The emissivities for the Stokes parameters are described by the fitting functions:
\begin{equation}
\begin{aligned}
I_I(x) & =2.5651\left(1+1.92 x^{-1 / 3}+0.9977 x^{-2 / 3}\right) \mathrm{e}^{-1.8899 x^{1 / 3}}\,, \\
I_Q(x) & =2.5651\left(1+0.93193 x^{-1 / 3}+0.499873 x^{-2 / 3}\right) \mathrm{e}^{-1.8899 x^{1 / 3}}\,, \\
I_V(x) & =\left(1.81348 x^{-1}+3.42319 x^{-2 / 3}+0.0292545 x^{-1 / 2}+2.03773 x^{-1 / 3}\right) \mathrm{e}^{-1.8899 x^{1 / 3}}\,.
\end{aligned}
\end{equation}

The absorption coefficients are obtained from the emission coefficients under the assumption of local thermodynamic equilibrium, so that Kirchhoff’s law, $\alpha_\nu=j_\nu/B_\nu$, holds with $B_\nu$ the blackbody function. The Faraday rotation coefficients, which affect the generation and propagation of polarization in a magnetized plasma, are given by
\begin{equation}
\begin{aligned}
& \rho_Q=\frac{n_e e^4 B^2 \sin ^2 \theta_B}{4 \pi^2 m_e^3 c^3 \nu^3} f_m(y)+\left(\frac{K_1\left(\Theta_e^{-1}\right)}{K_2\left(\Theta_e^{-1}\right)}+6 \Theta_e\right), \\
& \rho_V=\frac{n_e e^3 B \cos \theta_B}{\pi m_e^2 c^2 \nu^2} \frac{K_0\left(\Theta_e^{-1}\right)-\Delta J_5(y)}{K_2\left(\Theta_e^{-1}\right)}\,,
\end{aligned}
\end{equation}
with
\begin{equation}
y=\left(\frac{3}{2 \sqrt{2}} \times 10^{-3} \frac{\nu}{\nu_c}\right)^{-1 / 2}\,.
\end{equation}
Here, $f_m$ and $\Delta J_5$ are fitting functions, 
\begin{equation}
\begin{aligned}
f_m(y)= & 2.011 \exp \left(-\frac{y^{1.035}}{4.7}\right)-\cos \left(\frac{y}{2}\right) \exp \left(-\frac{y^{1.2}}{2.73}\right)-0.011 \exp \left(-\frac{y}{47.2}\right) \\
& +\frac{1}{2}\left[0.011 \exp \left(-\frac{y}{47.2}\right)-2^{-1 / 3} 3^{-23 / 6} \pi \times 10^4 y^{-8 / 3}\left(1+\tanh \left(10 \ln \frac{y}{120}\right)\right)\right]\,,\\
\Delta J_5(y)&=0.4379 \ln \left(1+0.001858 y^{1.503}\right)\,.
\end{aligned}
\end{equation}
and $K_n(y)$ denotes the modified Bessel function of the second kind of order $n$.

The detailed expressions for the emission, absorption, and Faraday rotation coefficients can be found in~\cite{Wang:2025gbj,Huang:2024bar}. In general, these coefficients primarily depend on $n_e$, $T_e$, $B$ and $\theta_B$.

We now turn to the specific intensity of the Kerr-MOG black hole image illuminated by the accretion disk. Along a complete light ray that connects the emitting plasma in the disk to the observer’s screen in the ZAMO frame~\cite{Hu:2020usx}, the light rays interact with matter in radiative transfer. This study employs the tensor form of the radiative transfer equation as presented in~\cite{2012ApJ...752..123G}:
\begin{equation}
\label{rteq}
k^\mu \nabla_\mu \mathcal{N}^{\alpha \beta}=\mathcal{J}^{\alpha \beta}+\mathcal{H}^{\alpha \beta \mu \nu} \mathcal{N}_{\mu \nu}\,,
\end{equation}
where $k^\mu$ is the photon wave vector, $\mathcal{N}^{\alpha \beta}$ is the polarization tensor, $\mathcal{J}^{\alpha \beta}$ represents emissivity of the source, and $\mathcal{H}^{\alpha \beta \mu \nu}$ characterizes absorption and Faraday rotation effects~\cite{Huang:2024bar}. Owing to the gauge invariance of $\mathcal{N}^{\alpha\beta}$, Eq.~\eqref{rteq} can be decomposed into two contributions in a parallel-transported frame~\cite{Huang}. The first part accounts for gravitational effects:
\begin{equation}
    k^\mu \Delta_\mu f^\nu=0\,,\,\,f^\nu k_\nu=0\,,
\end{equation}
where $f^\mu$ is a normalized spacelike vector orthogonal to $k^\mu$. The second part describes the interaction with the magnetized plasma:
\begin{equation}
\frac{\df}{\df \lambda} \mathcal{S}=R(\vartheta) J-R(\vartheta) \mathcal{M} R(-\vartheta) \mathcal{S}\,,
\end{equation}
where
\begin{equation}
\mathcal{S}=\left(\begin{array}{l}
\mathcal{I} \\
\mathcal{Q} \\
\mathcal{U} \\
\mathcal{V}
\end{array}\right), \quad J=\frac{1}{\nu^2}\left(\begin{array}{l}
j_I \\
j_Q \\
j_U \\
j_V
\end{array}\right), \quad \mathcal{M}=\nu\left(\begin{array}{cccc}
\alpha_I & \alpha_Q & \alpha_U & \alpha_V \\
\alpha_Q & \alpha_I & \rho_V & -\rho_U \\
\alpha_U & -\rho_V & \alpha_I & \rho_Q \\
\alpha_V & \rho_U & -\rho_Q & \alpha_I
\end{array}\right) \,.
\end{equation}
Here, $\{ \mathcal{I}, \mathcal{Q}, \mathcal{U}, \mathcal{V} \}
= \{ I/\nu^3,\, Q/\nu^3,\, U/\nu^3,\, V/\nu^3 \}$ represent the invariant Stokes parameters. The rotation matrix 
\begin{equation}
\label{rotationmatrix}
R(\vartheta)=\left(\begin{array}{cccc}
1 & & & \\
& \cos (2 \vartheta) & -\sin (2 \vartheta) & \\
& \sin (2 \vartheta) & \cos (2 \vartheta) & \\
& & & 1
\end{array}\right)\,,
\end{equation}
represents the rotation between the synchrotron emission frame and the parallel-transported reference frame. The rotation angle $\vartheta$ is defined as the angle between the reference vector $f^\mu$ and the magnetic field $B^\mu$ in the transverse subspace orthogonal to $k^\mu$:
\begin{equation}
\vartheta=\operatorname{sign}\left(\epsilon_{\mu \nu \alpha \beta} u^\mu f^\nu B^\rho k^\sigma\right) \arccos \left(\frac{h^{\mu \nu} f_\mu B_\nu}{\sqrt{\left(h^{\mu \nu} f_\mu f_\nu\right)\left(h^{\alpha \beta} B_\alpha B_\beta\right)}}\right)\,,
\end{equation}
where $h^{\mu\nu}=g^{\mu\nu}-\frac{k^\mu k^\nu}{\omega^2}+\frac{u^\mu k^\nu+k^\mu u^\nu}{\omega}$ is the induced metric on the transverse subspace.

To obtain the Stokes parameters on the observer’s screen, we apply the rotation matrix of~\eqref{rotationmatrix}. The relevant rotation angle $\vartheta_o$ is between the reference vector $f^\mu$ and $Y$-axis direction on the screen, which is taken as $-\partial_\theta^\mu$ in the transverse subspace orthogonal to $k^\mu$. The resulting projected Stokes parameters are
\begin{equation}
I_o=I, \quad Q_o=Q \cos 2\vartheta_o-U \sin 2\vartheta_o, \quad U_o=Q \sin 2\vartheta_o+U \cos 2\vartheta_o, \quad V_o=V\,.
\end{equation}

The measured Stokes parameters contain the full information on the polarization state of the received radiation. Here, $I_o$ represents the total intensity, and $V_o$ measures circular polarization, with its sign distinguishing the handedness of circular polarization (positive for right-handed and negative for left-handed). The degree of linear polarization and electric-vector position angle (EVPA) are
\begin{equation}
P_o=\frac{\sqrt{Q_o^2+U_o^2}}{I_o}, \quad \chi=\frac{1}{2} \arctan \frac{U_o}{Q_o}\,.
\end{equation}

On the observer’s image plane, it is convenient to introduce a two-dimensional polar coordinate system defined by
\bea
X=\rho\cos\varphi\,,\quad Y=\rho\sin\varphi\,,
\eea
where $\rho$ denotes the radial distance from the image center and $\varphi$ is the azimuthal angle measured counterclockwise on the image plane. 

Next, we focus on black hole imaging under specific accretion flow models. In this work, we consider two representative cases: the phenomenological RIAF and the analytic BAAF model. For the RIAF, we examine the intensity maps of an infalling accretion flow (Eq.~\eqref{infalling}), considering both isotropic and anisotropic emission case. For the BAAF model, we study the intensity distribution and the corresponding polarization patterns under anisotropic radiation.

\section{RIAF Model}
\label{sec3}

Similar to the construction method in~\cite{2011ApJ...735..110B}, the density and temperature distribution profiles
can be defined as
\begin{equation}
n_e=n_h\left(\frac{r_h}{r}\right)^{2} \exp \left(-\frac{z^2}{2 R^2}\right), \quad T_e=T_h\left(\frac{r_h}{r}\right)\,,
\end{equation}
where the subscript ``$h$'' denotes quantities evaluated at the event horizon. In all of these, $R = r \sin\theta$ is the cylindrical radius and $z = r \cos\theta$ measures the vertical height from $\theta=\pi/2$.

The local magnetic field strength is parameterized by a cold magnetization factor $\kappa$, defined as $B = (\kappa\,\rho_{\rm fluid})^{1/2}$, where $\rho_{\rm fluid} = n_e m_p c^2$ denotes the fluid mass density. Throughout this work we adopt a representative value $\kappa = 0.1$, consistent with the typical magnitude used in~\cite{Pu:2016qak}. We consider the infalling motion given by Eq.~\eqref{infalling}. Assuming ideal MHD, $u_{\mu}F^{\mu\nu}=0$, we adopt a purely toroidal magnetic-field configuration,
\begin{equation}
\label{toroidalmag}
b^\mu \sim (0,\,0,\,0,\,1)\,.
\end{equation}

For isotropic radiation, where the emissivity is independent of the pitch angle $\theta_B$, we employ an angle-averaged emissivity defined as  
\begin{equation}
\bar{j}_I=\frac{1}{2} \int_0^\pi j_I \sin \theta_B \mathrm{d}\theta_B\,,
\end{equation}
with the fitting formula from ~\cite{Mahadevan:1996cc}:
\begin{equation}
\bar{j}_I=\frac{n_e e^2 \nu}{2 \sqrt{3} c \Theta_e^2} M_I(x^\prime), \quad x^\prime=\frac{\nu}{\nu_c^\prime}, \quad \nu_c^\prime=\frac{3 e B \Theta_e^2}{4 \pi m_e c}\,,
\end{equation}
where the dimensionless function $M_I(x^\prime)$ is approximated as~\cite{Mahadevan:1996cc}
\begin{equation}
M_I(x^\prime)=\frac{4.0505}{{x^\prime}^{1 / 6}}\left(1+\frac{0.4}{{x^\prime}^{1 / 4}}+\frac{0.5316}{{x^\prime}^{1 / 2}}\right) \exp \left(-1.8899 {x^\prime}^{1 / 3}\right)\,.
\end{equation}

\subsection{Isotropic Radiation Case}

It should be emphasized that synchrotron radiation is intrinsically anisotropic, with its emissivity strongly dependent on the emission direction.
This dependence is governed by the pitch angle $\theta_B$, defined as the angle between the photon wave vector and the magnetic field in the fluid rest frame. To study the influence of such anisotropy on the black hole images, it is necessary to consider an isotropic emission profile as a comparison.

\begin{figure*}[htbp]
  \centering
  \begin{subfigure}{0.23\textwidth}
    \includegraphics[width=3.5cm,height=3cm]{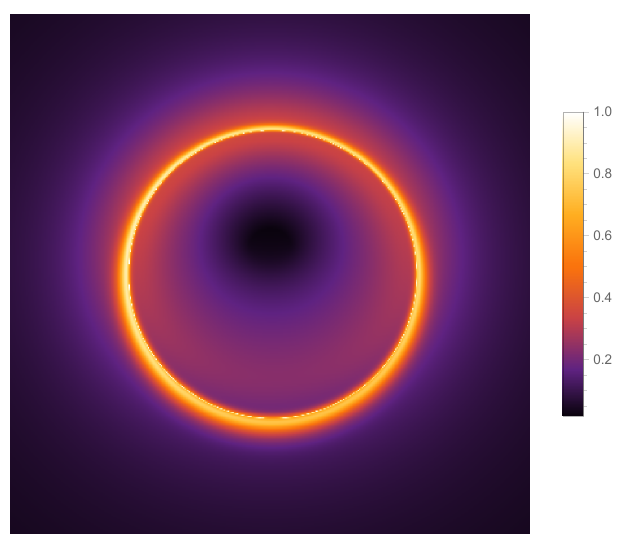}
    \caption{$a=0.1,\alpha=0.2$}
  \end{subfigure}
  \begin{subfigure}{0.23\textwidth}
    \includegraphics[width=3.5cm,height=3cm]{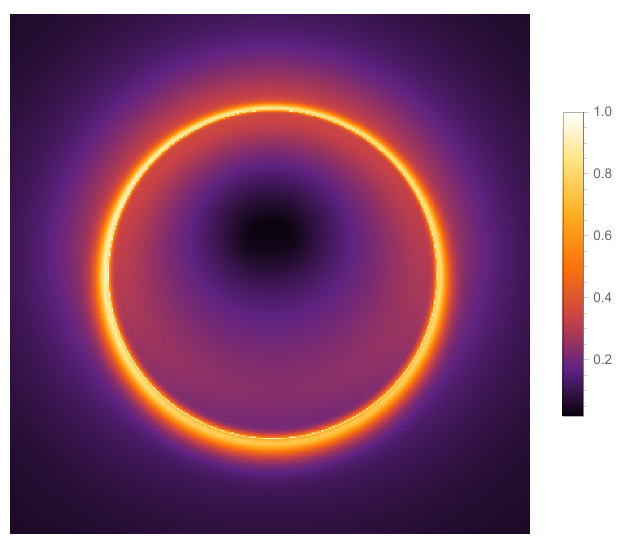}
   \caption{$a=0.1,\alpha=0.4$}
  \end{subfigure}
  \begin{subfigure}{0.23\textwidth}
    \includegraphics[width=3.5cm,height=3cm]{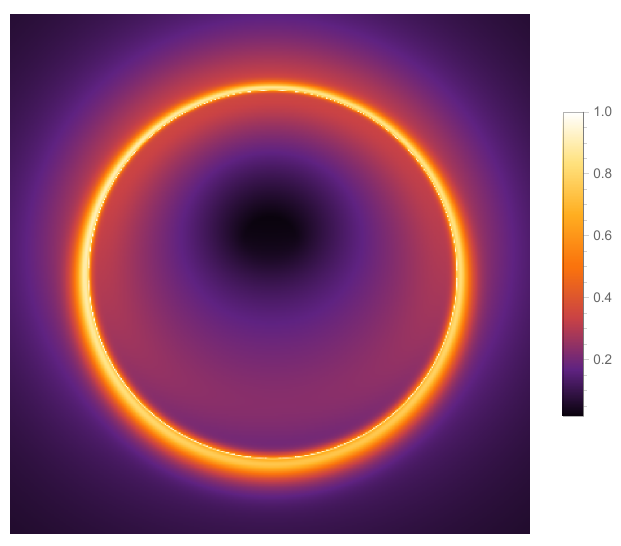}
  \caption{$a=0.1,\alpha=0.6$}
  \end{subfigure}
  \begin{subfigure}{0.23\textwidth}
    \includegraphics[width=3.5cm,height=3cm]{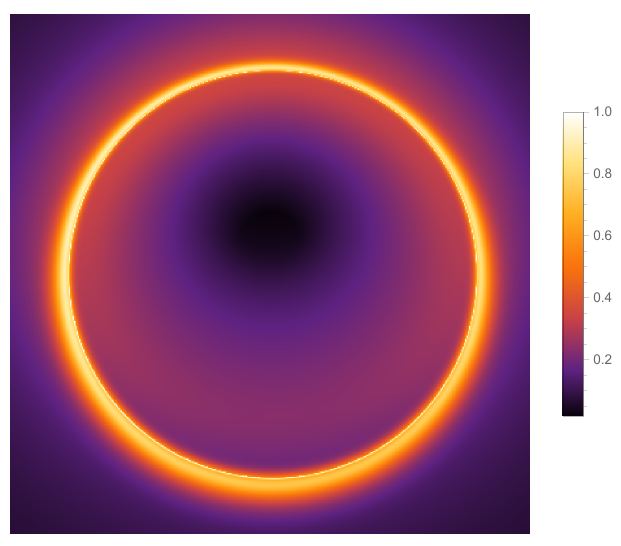}
    \caption{$a=0.1,\alpha=0.8$}
  \end{subfigure}
 \begin{subfigure}{0.23\textwidth}
    \includegraphics[width=3.5cm,height=3cm]{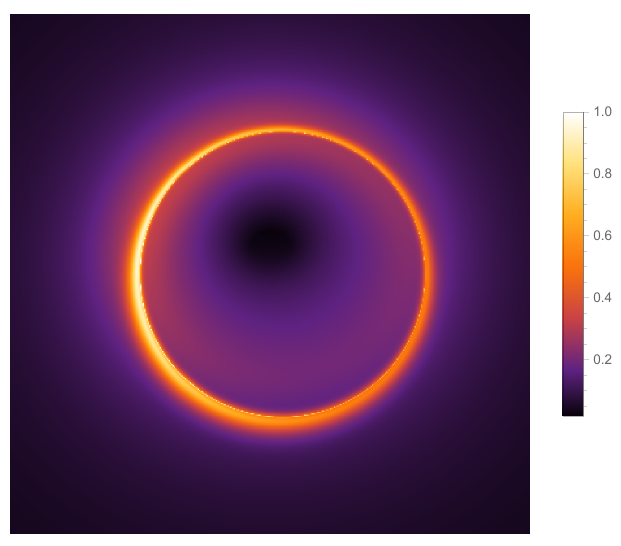}
    \caption{$a=0.5,\alpha=0.2$}
  \end{subfigure}
  \begin{subfigure}{0.23\textwidth}
    \includegraphics[width=3.5cm,height=3cm]{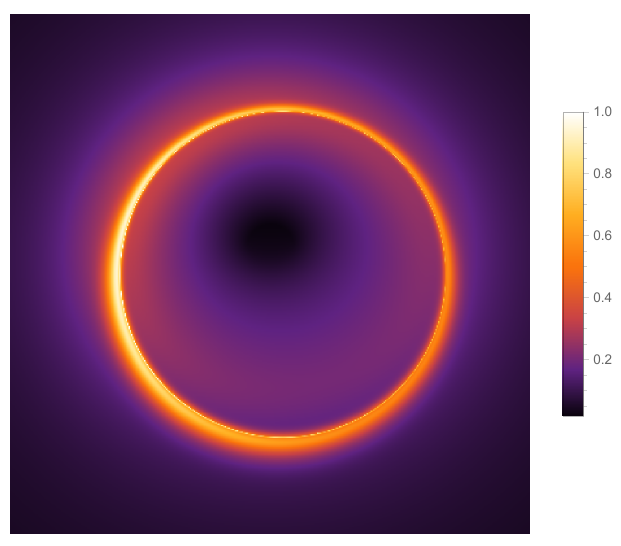}
    \caption{$a=0.5,\alpha=0.4$}
  \end{subfigure}
   \begin{subfigure}{0.23\textwidth}
    \includegraphics[width=3.5cm,height=3cm]{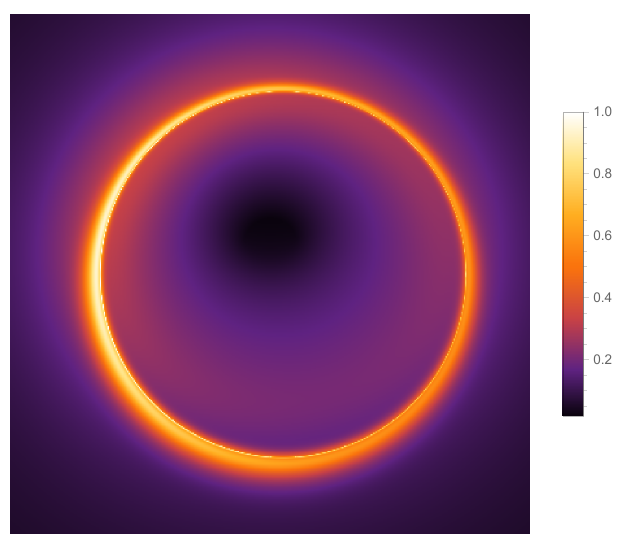}
    \caption{$a=0.5,\alpha=0.6$}
  \end{subfigure}
  \begin{subfigure}{0.23\textwidth}
    \includegraphics[width=3.5cm,height=3cm]{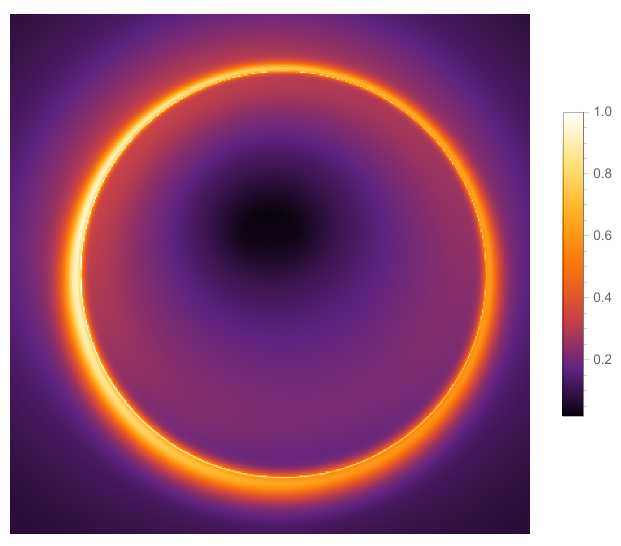}
    \caption{$a=0.5,\alpha=0.8$}
  \end{subfigure}
  \begin{subfigure}{0.23\textwidth}
    \includegraphics[width=3.5cm,height=3cm]{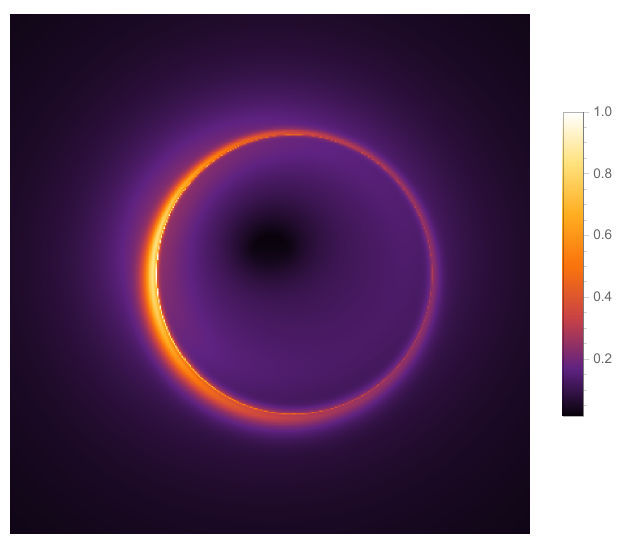}
    \caption{$a=0.9,\alpha=0.2$}
  \end{subfigure}
  \begin{subfigure}{0.23\textwidth}
    \includegraphics[width=3.5cm,height=3cm]{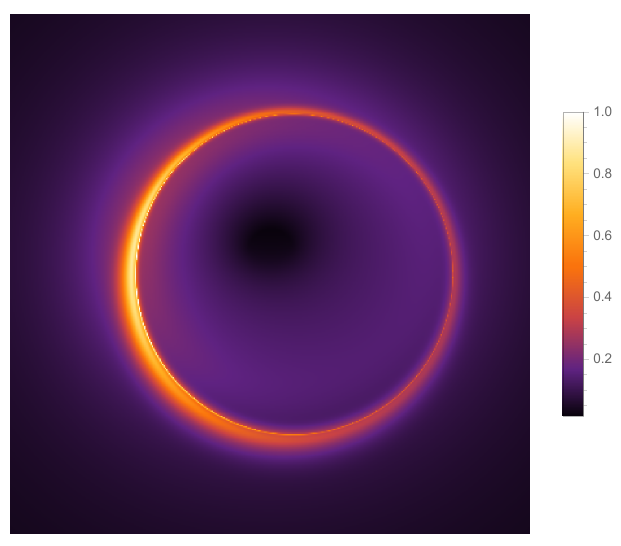}
    \caption{$a=0.9,\alpha=0.4$}
  \end{subfigure}
  \begin{subfigure}{0.23\textwidth}
    \includegraphics[width=3.5cm,height=3cm]{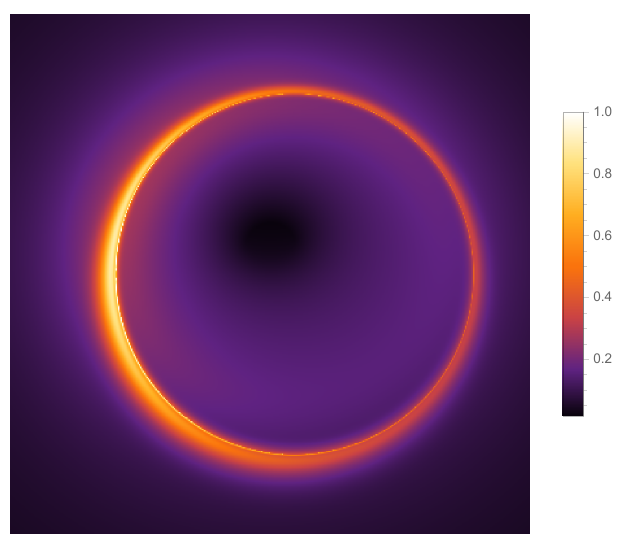}
    \caption{$a=0.9,\alpha=0.6$}
  \end{subfigure}
  \begin{subfigure}{0.23\textwidth}
    \includegraphics[width=3.5cm,height=3cm]{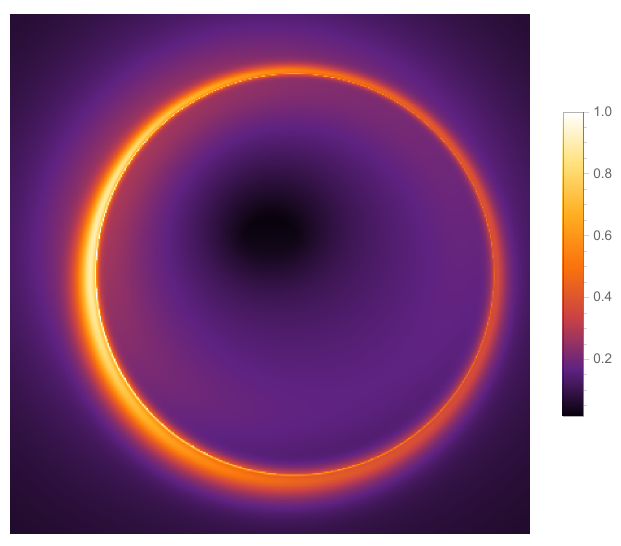}
    \caption{$a=0.9,\alpha=0.8$}
  \end{subfigure}
   \caption{Intensity maps in the RIAF model with isotropic emission. 
    The motion of the accretion flow corresponds to the infalling case. 
    From left to right, the spin parameter takes the values 
    $a = 0.1,~0.5,~0.9$; from top to bottom, the MOG parameter is set to 
    $\alpha = 0.2,~0.4,~0.6,~0.8$. 
    The observation is performed at a frequency of $345~\mathrm{GHz}$ and $\theta_o=30^\circ$.}
   \label{fig:1}
\end{figure*}
\begin{figure*}[htbp]
  \centering
  \begin{subfigure}{0.45\textwidth}
    \includegraphics[width=6.5cm,height=3.5cm]{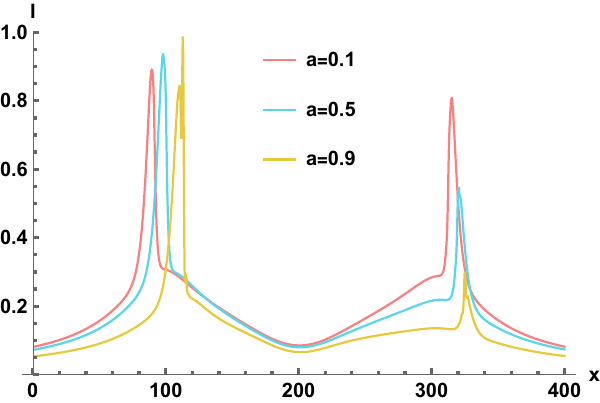}
    \caption{Horizontal}
  \end{subfigure}
 \begin{subfigure}{0.45\textwidth}
    \includegraphics[width=6.5cm,height=3.5cm]{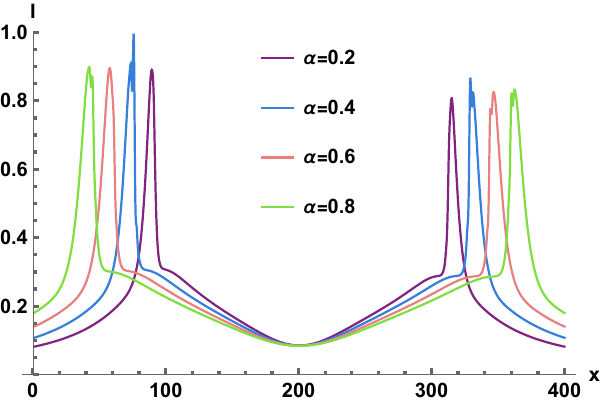}
    \caption{Horizontal}
  \end{subfigure}\\
  \begin{subfigure}{0.45\textwidth}
    \includegraphics[width=6.5cm,height=3.5cm]{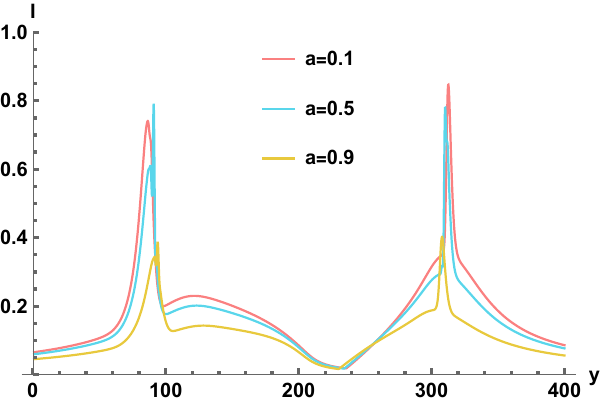}
    \caption{Vertical}
  \end{subfigure}
  \begin{subfigure}{0.45\textwidth}
    \includegraphics[width=6.5cm,height=3.5cm]{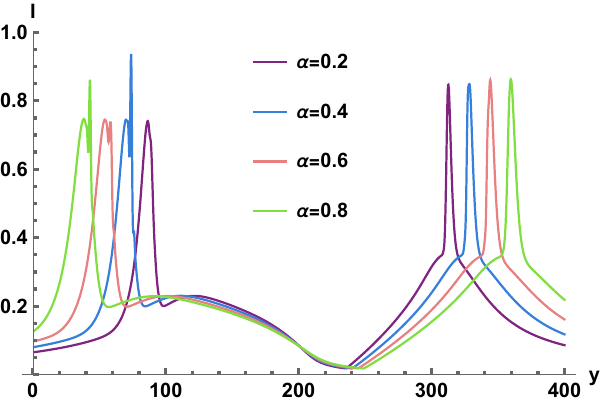}
    \caption{Vertical}
  \end{subfigure}
   \caption{Intensity distribution for the Kerr-MOG black hole in the RIAF model with isotropic emission. 
     The accretion flow follows the infalling motion.
     The observation is performed at a frequency of $345~\mathrm{GHz}$ and $\theta_o=30^\circ$.
    Left column: profiles for varying spin $a$ with $\alpha=0.2$. 
    Right column: profiles for varying deformation $\alpha$ with $a=0.1$. 
    Upper and lower panels correspond to horizontal ($X$-axis) and vertical ($Y$-axis) cuts, respectively.}
   \label{fig:2}
\end{figure*}
We present in Fig.~\ref{fig:1} the intensity maps for different spin parameters $a$ and MOG deformation parameters $\alpha$. The accretion flow is modeled using the RIAF prescription with an infalling motion and isotropic emission, while the observer’s inclination is fixed at $30^\circ$.
For a more quantitative analysis, Figure~\ref{fig:2} shows the corresponding intensity distributions along the $X$- and $Y$-axes of the observer’s screen.
All panels in Fig.~\ref{fig:1} display a prominent bright ring, corresponding to the intensity peaks shown in Fig.~\ref{fig:2}. Within this bright ring, a central dark region can be seen, which coincides with the near-zero intensity area in Fig.~\ref{fig:2}. The bright ring arises from higher-order images caused by strong gravitational lensing, while the surrounding diffuse region corresponds to the primary image. The dark region is associated with the event horizon. For geometrically thin disks, this region exhibits a sharp boundary, often referred to as the inner shadow that could be detected by the EHT~\cite{Chael:2021rjo}. In contrast, for geometrically thick disks, emission from off-equatorial regions partially obscures the horizon, making the inner shadow less distinct.
Comparing the columns in Fig.~\ref{fig:1} and the left panels in Fig.~\ref{fig:2}, we find that for fixed $\alpha$, both the bright ring and the central dark region shrink with increasing spin $a$.
Moreover, due to the frame-dragging effect induced by rotation, the intensity on the left side of the image becomes significantly enhanced as $a$ increases.
In contrast, by examining the rows in Fig.~\ref{fig:1} and the right panels in Fig.~\ref{fig:2}, we observe that for fixed spin $a$, both the bright ring and the central dark region expand with increasing $\alpha$, and the ring width also slightly broadens.

\begin{figure*}[htbp]
  \centering
  \begin{subfigure}{0.23\textwidth}
    \includegraphics[width=3.5cm,height=3cm]{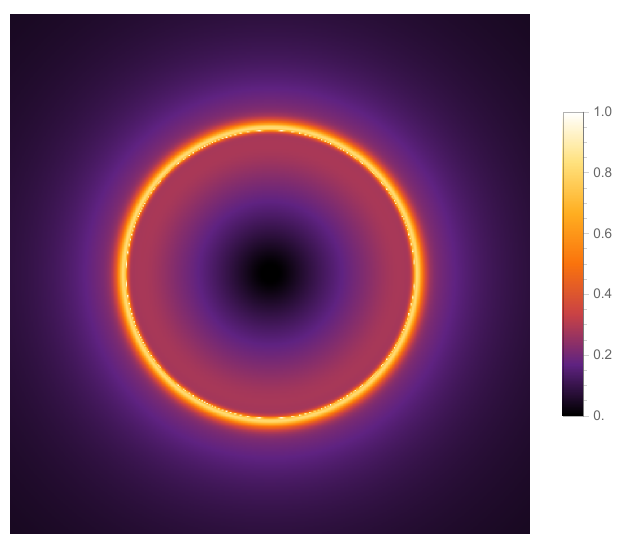}
    \caption{$\theta_o=1^\circ$}
  \end{subfigure}
  \begin{subfigure}{0.23\textwidth}
    \includegraphics[width=3.5cm,height=3cm]{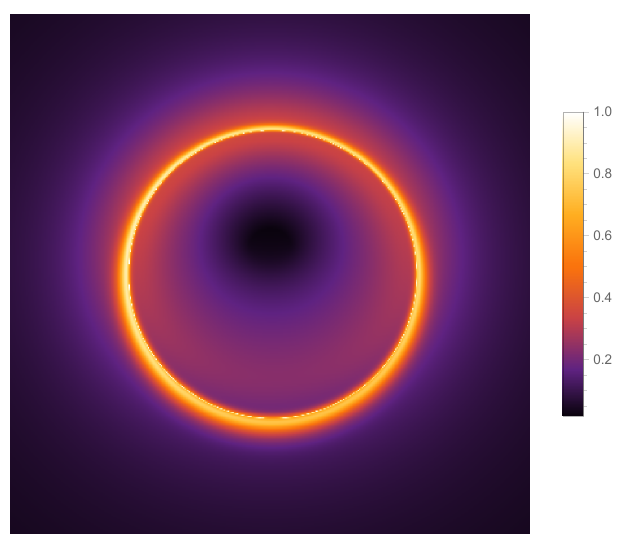}
   \caption{$\theta_o=30^\circ$}
  \end{subfigure}
  \begin{subfigure}{0.23\textwidth}
    \includegraphics[width=3.5cm,height=3cm]{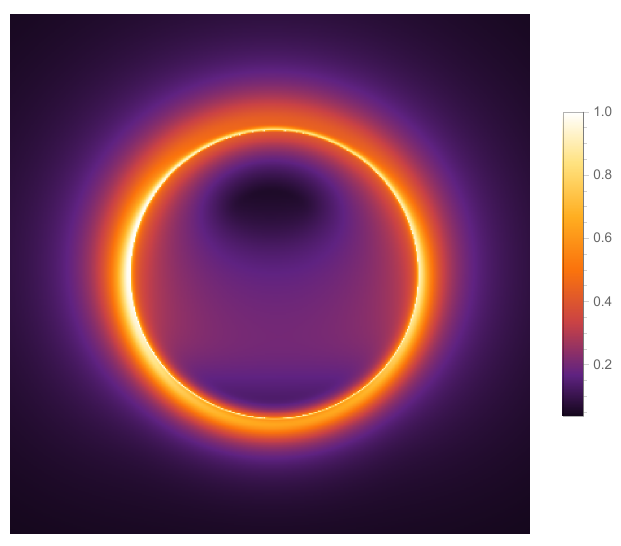}
   \caption{$\theta_o=60^\circ$}
  \end{subfigure}
  \begin{subfigure}{0.23\textwidth}
    \includegraphics[width=3.5cm,height=3cm]{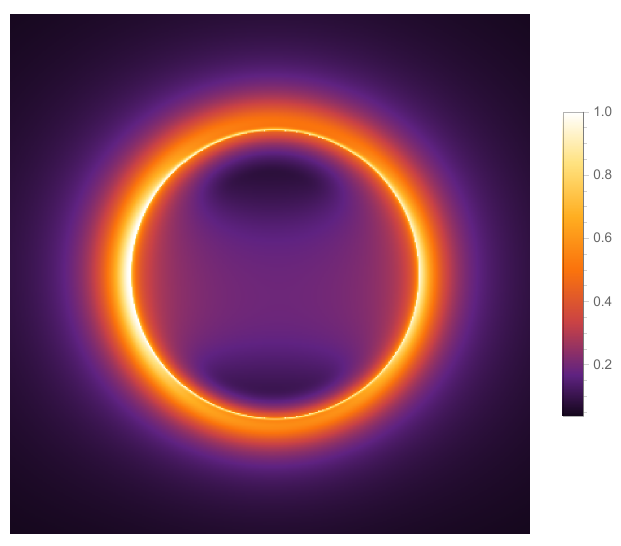}
    \caption{$\theta_o=80^\circ$}
  \end{subfigure}
   \caption{ Intensity maps in the RIAF model with isotropic emission 
    for different inclinations.
    The motion of the accretion flow corresponds to the infalling case.
    From left to right, $\theta_o = 1^\circ,~30^\circ,~60^\circ,~80^\circ$.
    The observation frequency is $345~\mathrm{GHz}$, with spin parameter $a = 0.1$ and MOG parameter $\alpha = 0.2$.}
   \label{fig:3}
\end{figure*}
In Fig.~\ref{fig:3}, we present the dependence of the intensity distribution on the observer’s inclination angle, keeping the spin $a$ and MOG parameter $\alpha$ fixed.
As the inclination increases, the image morphology undergoes noticeable changes. Owing to the nearly spherical symmetry of the spacetime ($a=0.1$) and the infalling motion of the accretion flow, the intensity map remains left–right symmetric.
For polar viewing, the bright ring (higher-order images) and the dark region remain centered and isotropic. For observers at 
$\theta_o=1^\circ\,,30^\circ$, the dark region remains clearly visible. However, at higher inclinations, 
$\theta_o=60^\circ\,,80^\circ$, two distinct black areas emerge, with the upper one appearing slightly darker than the lower. That's because for observers close to the equatorial plane, high-latitude emission partially fills the central darkness, whereas for observers near the poles, fewer photons are able to reach the line of sight, resulting in a more pronounced dark region.

\begin{figure*}[htbp]
  \centering
  \begin{subfigure}{0.3\textwidth}
    \includegraphics[width=4.8cm,height=4.2cm]{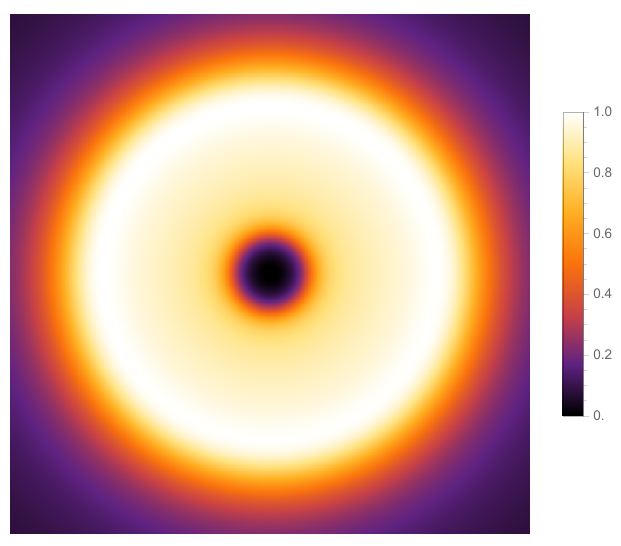}
    \caption{85~GHz}
  \end{subfigure}
  \begin{subfigure}{0.3\textwidth}
    \includegraphics[width=4.8cm,height=4.2cm]{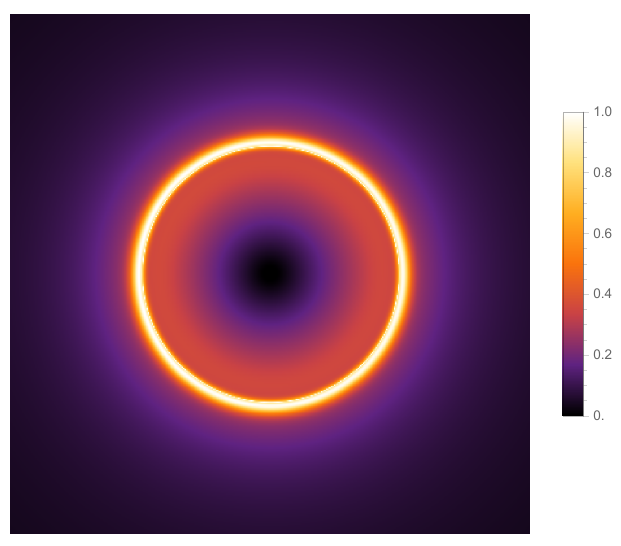}
    \caption{230~GHz}
  \end{subfigure}
  \begin{subfigure}{0.3\textwidth}
    \includegraphics[width=4.8cm,height=4.2cm]{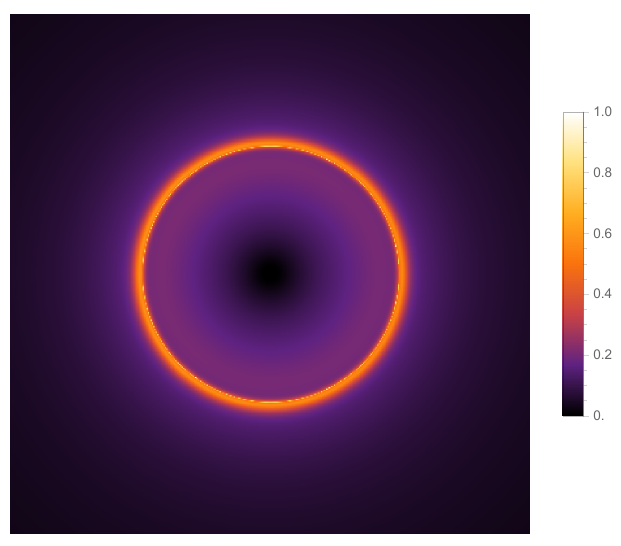}
    \caption{345~GHz}
  \end{subfigure}\\
  \begin{subfigure}{0.45\textwidth}
    \includegraphics[width=6.5cm,height=3.5cm]{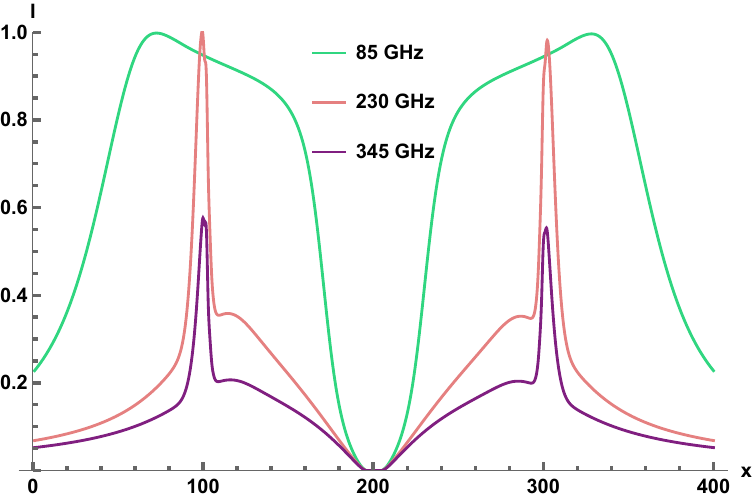}
    \caption{Horizontal}
  \end{subfigure}
  \begin{subfigure}{0.45\textwidth}
    \includegraphics[width=6.5cm,height=3.5cm]{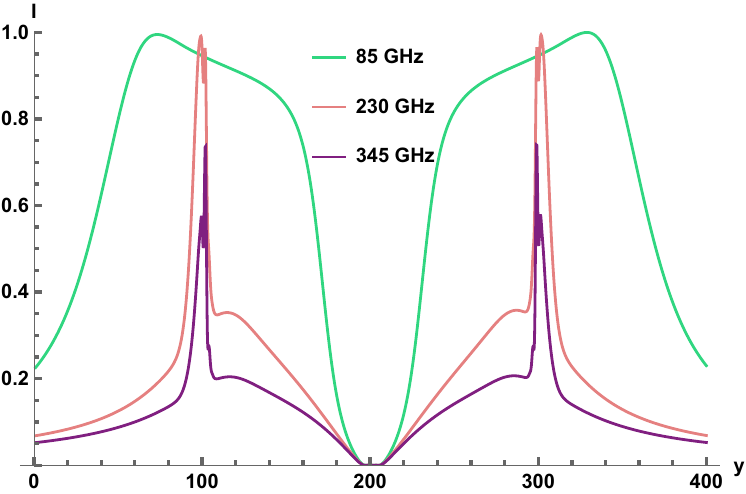}
    \caption{Vertical}
  \end{subfigure}
   \caption{Intensity maps in the RIAF with isotropic emission and infalling motion.
    The top row shows the images at observing frequencies of 85, 230, and 345~GHz (from left to right).
    The bottom row presents the horizontal (left) and vertical (right) intensity cuts for different frequencies. The inclination angle is fixed at $1^\circ$, with $a = 0.9$ and $\alpha = 0.6$.}
   \label{fig:4}
\end{figure*}

Figure \ref{fig:4} illustrates the observation frequency dependence of the black hole images in the RIAF model with isotropic emission and infalling motion.
At low observation frequency ($85~\mathrm{GHz}$), the image exhibits a diffuse and extended morphology dominated by optically thick emission.
As the observation frequency increases to $230~\mathrm{GHz}$ and $345~\mathrm{GHz}$, the image transitions to a sharper and more compact ring structure.
This behavior arises from the reduction of optical depth and the dominance of emission from regions closer to the event horizon at higher frequencies.
The corresponding intensity profiles in the lower panels exhibit pronounced peaks associated with the bright ring, with the peak amplitude increasing and the profile width narrowing as the observation frequency increases.
These trends are consistent with the expectation that higher observation frequency radiation traces regions deeper within the gravitational potential well, revealing the characteristic photon ring structure predicted by general relativity.

\subsection{Anisotropic Radiation Case}

\begin{figure*}[htbp]
  \centering
  \begin{subfigure}{0.23\textwidth}
    \includegraphics[width=3.5cm,height=3cm]{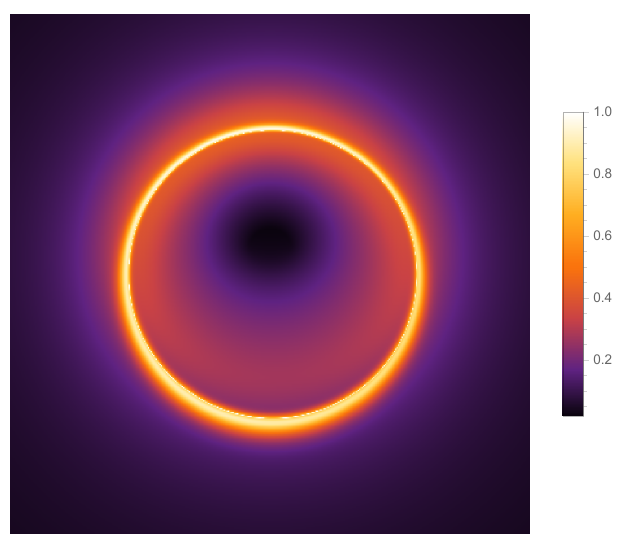}
    \caption{$a=0.1,\alpha=0.2$}
  \end{subfigure}
  \begin{subfigure}{0.23\textwidth}
    \includegraphics[width=3.5cm,height=3cm]{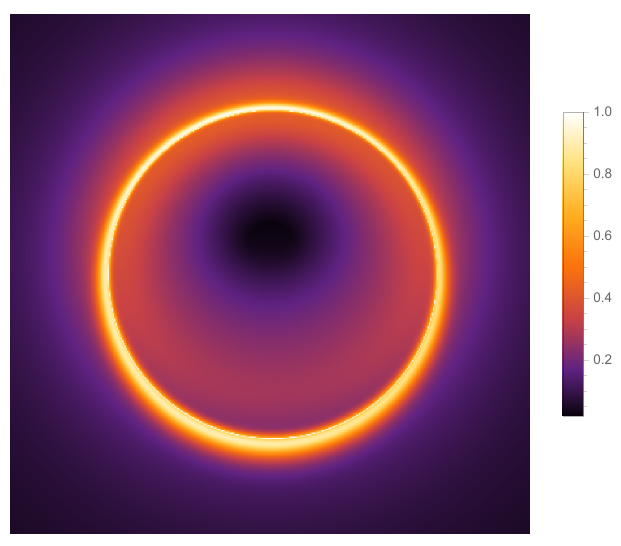}
   \caption{$a=0.1,\alpha=0.4$}
  \end{subfigure}
  \begin{subfigure}{0.23\textwidth}
    \includegraphics[width=3.5cm,height=3cm]{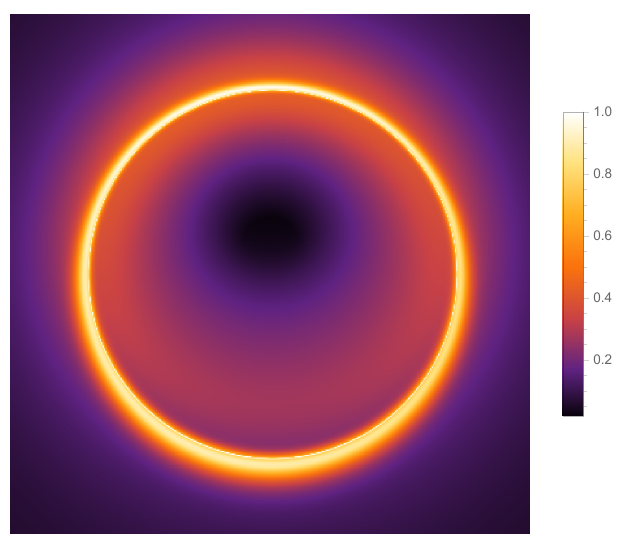}
  \caption{$a=0.1,\alpha=0.6$}
  \end{subfigure}
  \begin{subfigure}{0.23\textwidth}
    \includegraphics[width=3.5cm,height=3cm]{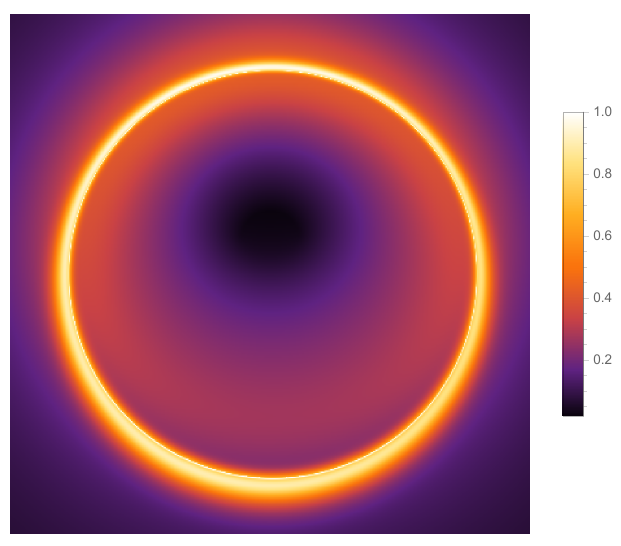}
    \caption{$a=0.1,\alpha=0.8$}
  \end{subfigure}
 \begin{subfigure}{0.23\textwidth}
    \includegraphics[width=3.5cm,height=3cm]{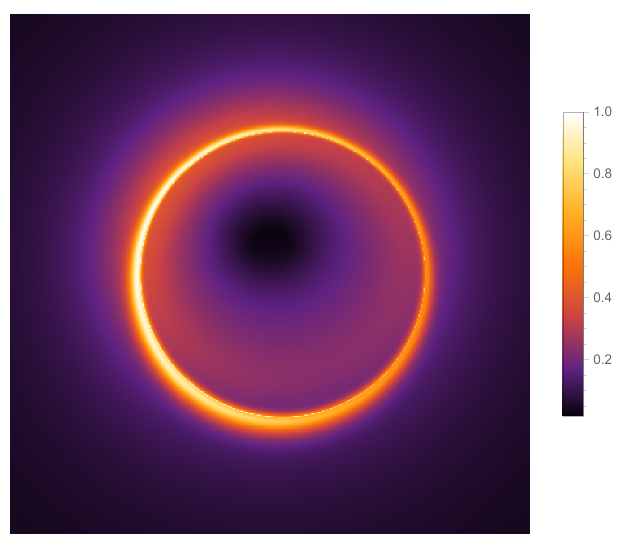}
    \caption{$a=0.5,\alpha=0.2$}
  \end{subfigure}
  \begin{subfigure}{0.23\textwidth}
    \includegraphics[width=3.5cm,height=3cm]{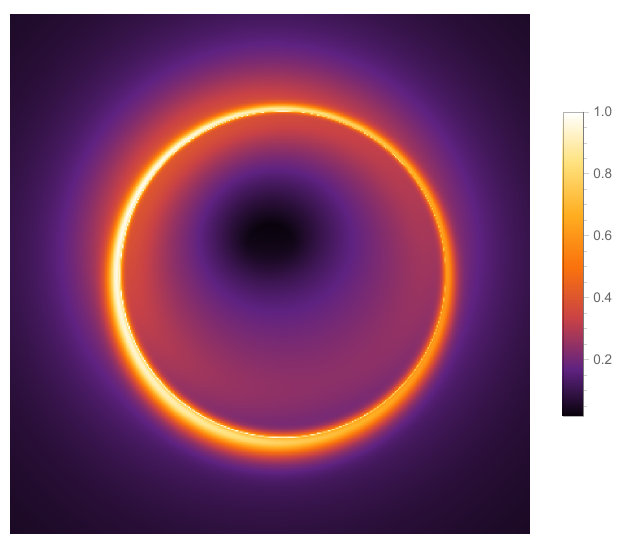}
    \caption{$a=0.5,\alpha=0.4$}
  \end{subfigure}
   \begin{subfigure}{0.23\textwidth}
    \includegraphics[width=3.5cm,height=3cm]{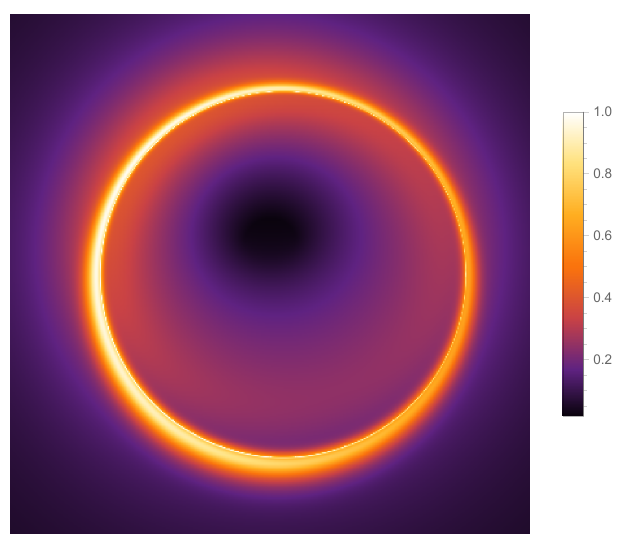}
    \caption{$a=0.5,\alpha=0.6$}
  \end{subfigure}
  \begin{subfigure}{0.23\textwidth}
    \includegraphics[width=3.5cm,height=3cm]{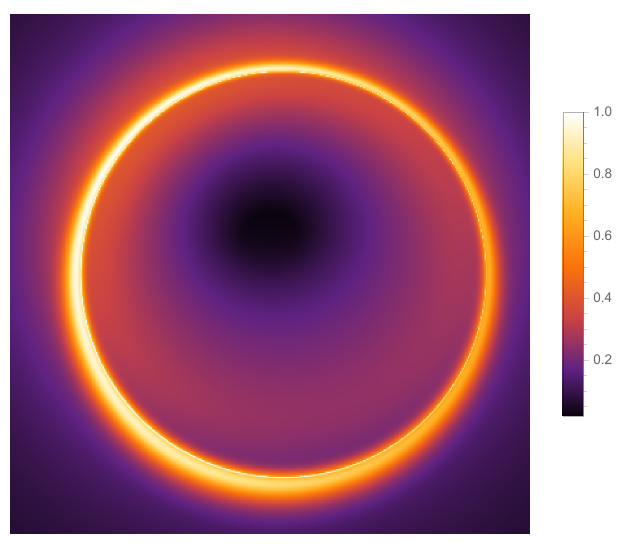}
    \caption{$a=0.5,\alpha=0.8$}
  \end{subfigure}
  \begin{subfigure}{0.23\textwidth}
    \includegraphics[width=3.5cm,height=3cm]{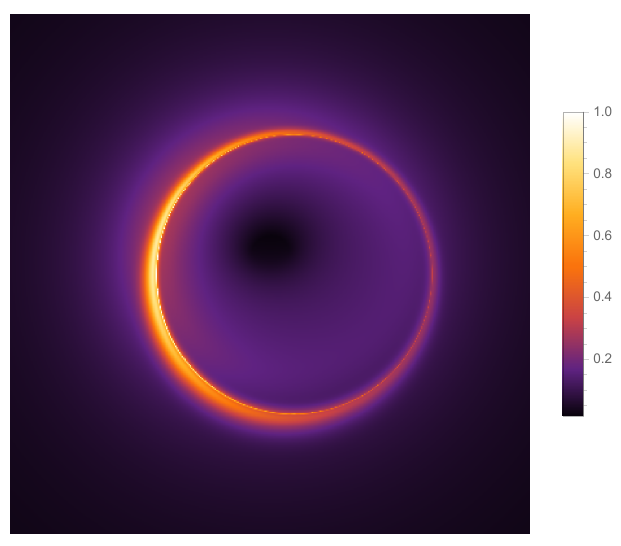}
    \caption{$a=0.9,\alpha=0.2$}
  \end{subfigure}
  \begin{subfigure}{0.23\textwidth}
    \includegraphics[width=3.5cm,height=3cm]{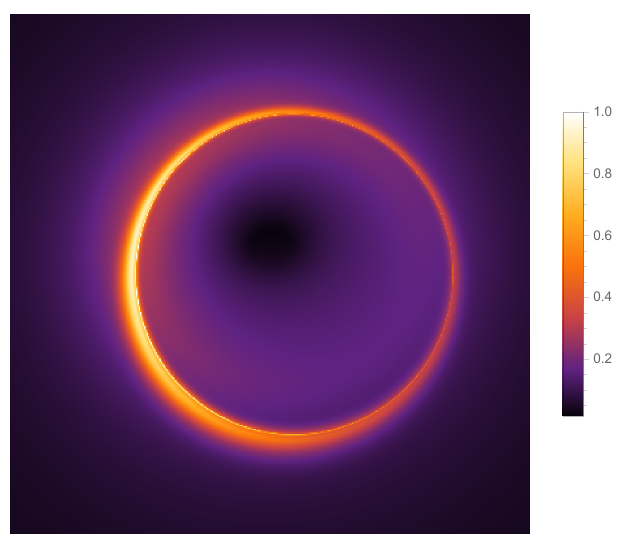}
    \caption{$a=0.9,\alpha=0.4$}
  \end{subfigure}
  \begin{subfigure}{0.23\textwidth}
    \includegraphics[width=3.5cm,height=3cm]{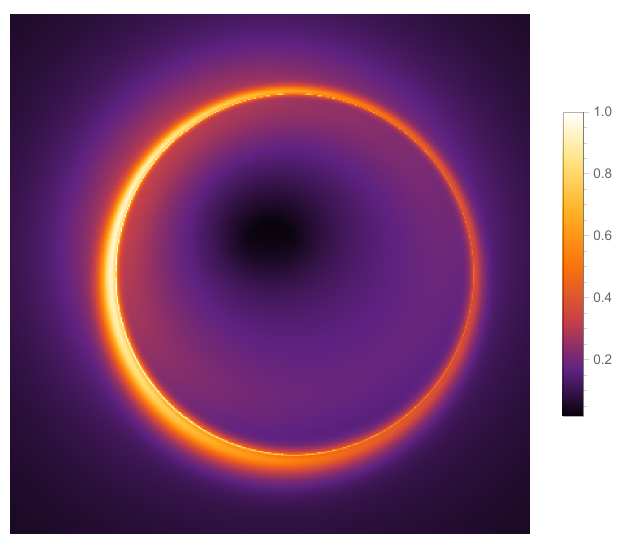}
    \caption{$a=0.9,\alpha=0.6$}
  \end{subfigure}
  \begin{subfigure}{0.23\textwidth}
    \includegraphics[width=3.5cm,height=3cm]{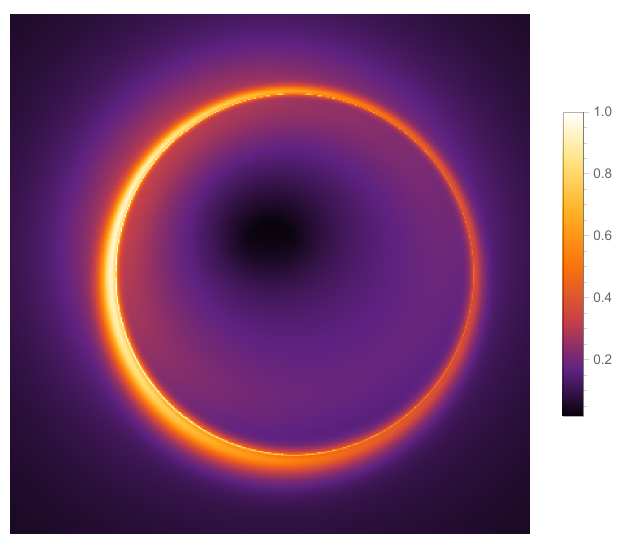}
    \caption{$a=0.9,\alpha=0.8$}
  \end{subfigure}
   \caption{ Intensity maps in the RIAF model with anisotropic emission. 
    The motion of the accretion flow corresponds to the infalling case. 
    From left to right, the spin parameter takes the values 
    $a = 0.1,~0.5,~0.9$; from top to bottom, the MOG parameter is set to 
    $\alpha = 0.2,~0.4,~0.6,~0.8$. 
    The observation frequency is $345~\mathrm{GHz}$, and the inclination angle is $30^\circ$.}
   \label{fig:5}
\end{figure*}

\begin{figure*}[htbp]
  \centering
  \begin{subfigure}{0.45\textwidth}
    \includegraphics[width=6.5cm,height=3.5cm]{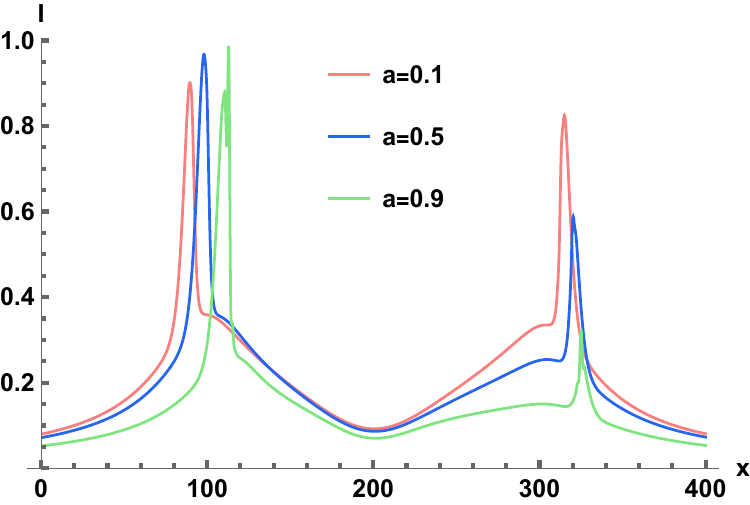}
    \caption{Horizontal}
  \end{subfigure}
  \begin{subfigure}{0.45\textwidth}
    \includegraphics[width=6.5cm,height=3.5cm]{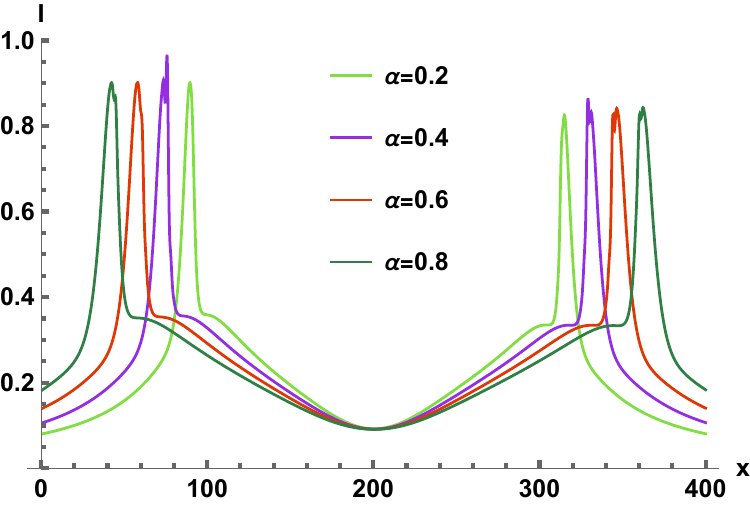}
    \caption{Horizontal}
  \end{subfigure}\\
    \begin{subfigure}{0.45\textwidth}
    \includegraphics[width=6.5cm,height=3.5cm]{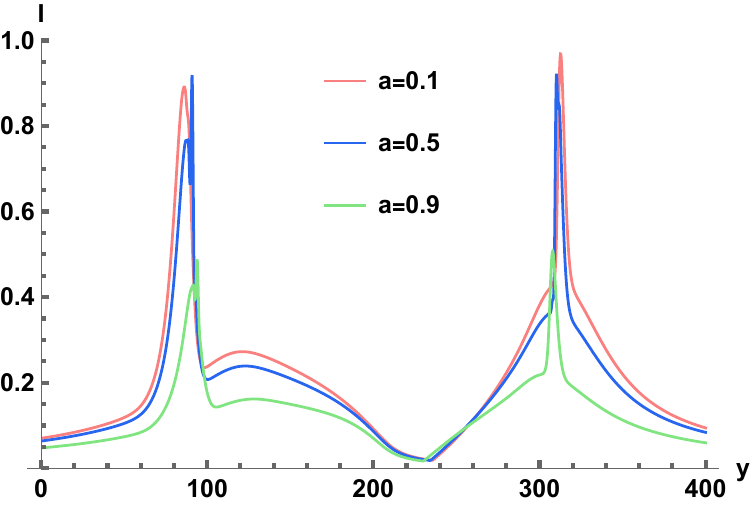}
    \caption{Vertical}
  \end{subfigure}
  \begin{subfigure}{0.45\textwidth}
    \includegraphics[width=6.5cm,height=3.5cm]{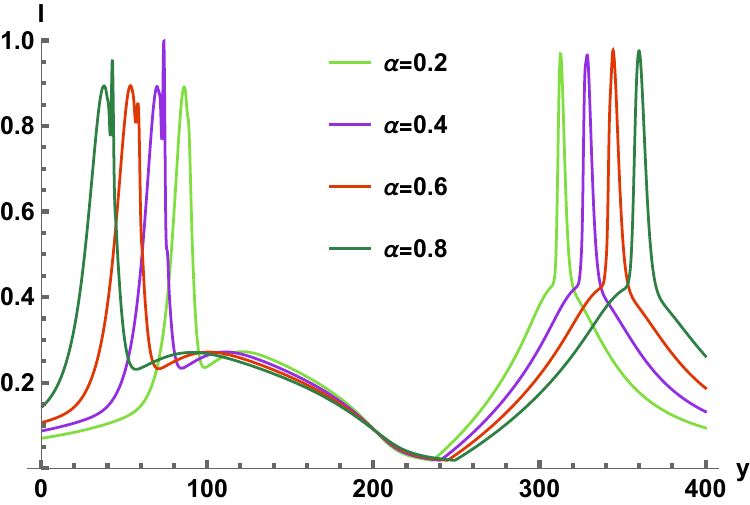}
    \caption{Vertical}
  \end{subfigure}
   \caption{Intensity distribution for the Kerr-MOG black hole in the RIAF model with anisotropic emission. 
     The accretion flow follows the infalling motion. The observer’s inclination is set to $30^\circ$.
    The left column shows profiles for varying spin $a$ with $\alpha=0.2$, 
    while the right column shows profiles for varying MOG parameter $\alpha$ with $a=0.1$.
    Upper and lower panels correspond to horizontal ($X$-axis) and vertical ($Y$-axis) cuts, respectively.}
   \label{fig:6}
\end{figure*}
Next, we extend our analysis to account for anisotropic synchrotron emission, assuming a toroidal magnetic field configuration as given in Eq.~\eqref{toroidalmag}.
Figure~\ref{fig:5} displays the corresponding intensity maps of the Kerr–MOG black hole within the RIAF model, for an observer inclination of $30^\circ$.
For quantitative comparison, the horizontal and vertical intensity profiles are shown in Fig.~\ref{fig:6}.

The overall morphology remains qualitatively similar to the isotropic case (Fig.~\ref{fig:1}), exhibiting a pronounced bright ring surrounding a central dark region, both of which expand with increasing $\alpha$.
As the spin parameter $a$ increases, the image becomes increasingly asymmetric, with enhanced brightness on the side corotating with the black hole due to frame dragging.

\begin{figure*}[htbp]
  \centering
  \begin{subfigure}{0.23\textwidth}
    \includegraphics[width=3.5cm,height=3cm]{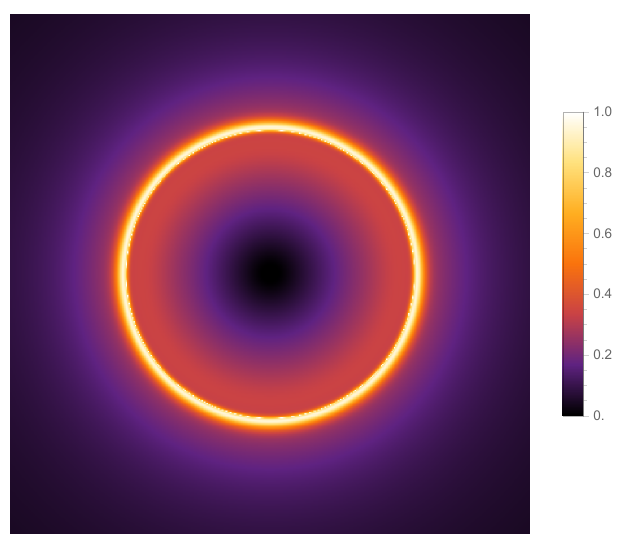}
    \caption{$\theta_o=1^\circ$}
  \end{subfigure}
  \begin{subfigure}{0.23\textwidth}
    \includegraphics[width=3.5cm,height=3cm]{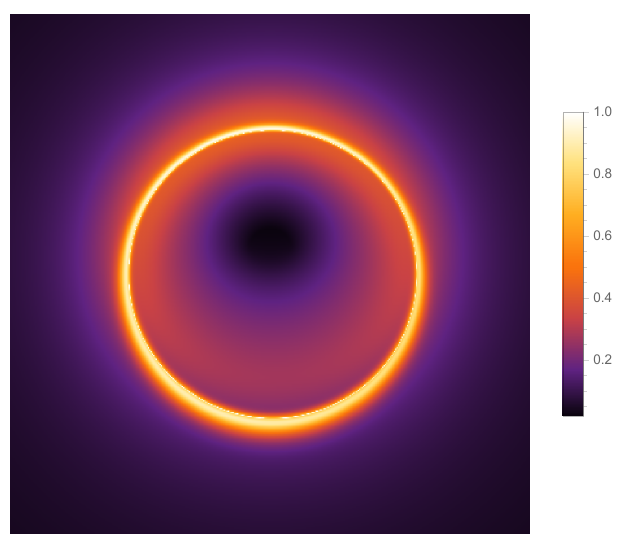}
   \caption{$\theta_o=30^\circ$}
  \end{subfigure}
  \begin{subfigure}{0.23\textwidth}
    \includegraphics[width=3.5cm,height=3cm]{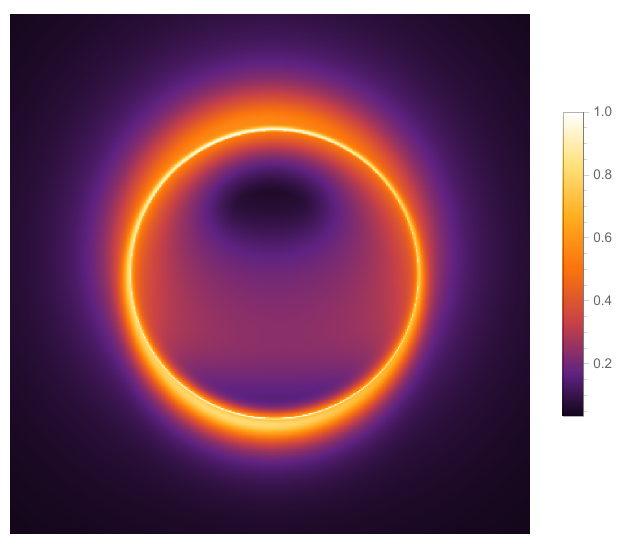}
   \caption{$\theta_o=60^\circ$}
  \end{subfigure}
  \begin{subfigure}{0.23\textwidth}
    \includegraphics[width=3.5cm,height=3cm]{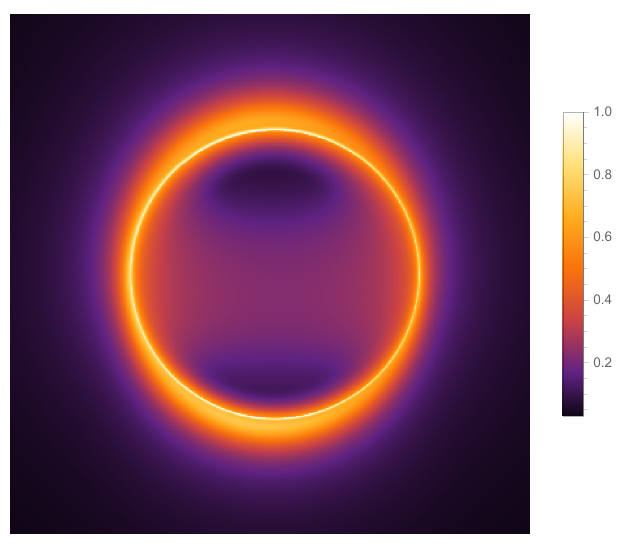}
    \caption{$\theta_o=80^\circ$}
  \end{subfigure}
   \caption{ Intensity maps in the RIAF model with anisotropic emission and infalling motion.
    Images are shown for observing inclinations $\theta_o = 1^\circ,\,30^\circ,\,60^\circ$, and $80^\circ$ (from left to right) at an observation frequency of $345~\mathrm{GHz}$.
    The model parameters are fixed to $a=0.1$ and $\alpha=0.2$.}
   \label{fig:7}
\end{figure*}
Figure~\ref{fig:7} illustrates the inclination dependence of the Kerr–MOG black hole images in the anisotropic emission model.
At higher inclinations, the brightness distribution becomes strongly nonuniform, with two distinct dark regions appearing inside the bright ring.
As the inclination increases, the intensity distribution becomes significantly asymmetric, and two distinct dark patches appear within the bright ring. At high inclination angles, the image morphology deviates markedly from the isotropic case: the bright ring becomes elliptical and is stretched along the vertical ($Y$) direction. This deformation results from the $\theta_B$–dependent nature of anisotropic synchrotron emission, where radiation from high-latitude plasma propagates nearly perpendicular to the magnetic field, enhancing the vertical emission and producing the elongated ring structure.

\section{BAAF Model}\label{sec4}

In the previous RIAF model, the magnetic field is modeled phenomenologically as a purely toroidal configuration, and thus the resulting polarization signatures do not fully capture the interplay between the black hole spacetime and the accretion dynamics.
To further elucidate the geometric and radiative characteristics of horizon-scale magnetofluids—particularly the polarization features—we now turn to a more analytically tractable BAAF model.

The BAAF disk is modeled as a steady, axisymmetric accretion flow in which the fluid has no motion in the polar direction, i.e., $u^\theta \equiv 0$, corresponding to a conical solution described in Eq.~\eqref{signl}~\cite{Hou:2023bep}. Under these assumptions, the mass conservation equation takes the form
\begin{equation}
\frac{\mathrm{d}}{\mathrm{~d} r}\left(\sqrt{-g} \rho u^r\right)=0, \quad \Longrightarrow \rho=\rho_h \frac{\left.\sqrt{-g} u^r\right|_{r=r_h}}{\sqrt{-g} u^r}\,,
\end{equation}
where $\rho$ is the rest mass density. By projecting the energy–momentum conservation equation $\nabla_\mu T^{\mu\nu}=0$ along the $u^\mu$, we obtain
\begin{equation}
\label{energy}
\df U=\frac{U+p}{\rho} \df \rho\,,
\end{equation}
where $p$ is the isotropic pressure, $U$ denotes the internal energy density of the fluid.  
Introducing the proton-to-electron temperature ratio $\epsilon = T_p/T_e$, we get
\begin{equation}
\label{xi}
U=\rho+\rho \frac{3}{2}(\epsilon+2) \frac{m_e}{m_p} \Theta_e\,,
\end{equation} 
From the ideal gas equation of
state, we have
\begin{equation}
\label{gas}
p=n k_B\left(T_p+T_e\right)=\rho(1+\epsilon) \frac{m_e}{m_p} \Theta_e\,,
\end{equation}
Substituting Eqs.~\eqref{xi} and~\eqref{gas} into Eq.~\eqref{energy} and integrating gives
\begin{equation}
\Theta_e=\left(\Theta_e\right)_h\left(\frac{\rho}{\rho_h}\right)^{\frac{2(1+\epsilon)}{3(2+\epsilon)}}\,,
\end{equation}

For the conical solution described in Sec.~\ref{metric}, the rest-mass density and electron temperature take the analytic form
\begin{equation}
\begin{aligned}
    \rho(r,\theta) &=\rho(r_h,\theta)\sqrt{\frac{\mathcal{R}_\text{c}(r_h,\theta)}{\mathcal{R}_\text{c}(r,\theta)}} \,,\\
    \Theta_e(r,\theta) &=\Theta_e(r_h,\theta)\left[\frac{\mathcal{R}_\text{c}(r_h,\theta)}{\mathcal{R}_\text{c}(r,\theta)}\right]^{\frac{1+\epsilon}{3(2+\epsilon)}}\,.
\end{aligned}
\end{equation}

For simplicity, we model $\rho(r_h,\theta)$ using a Gaussian distribution and take $\Theta_e(r_h,\theta)$ to be uniform at the horizon:
\begin{equation}
\rho\left(r_h, \theta\right)=\rho_h \exp \left[-\left(\frac{\sin \theta-\sin \theta_\text{c}}{\sigma}\right)^2\right], \quad \Theta\left(r_h, \theta\right)=\Theta_h\,.
\end{equation}
Here, $\theta_\text{c}$  determines the central latitude and $\sigma$ sets the angular width.  
We adopt an equatorially symmetric thick disk with $\theta_\text{c}=\pi/2$ and $\sigma=1/5$. Throughout this work, we set the boundary values
$\rho_h \simeq 1.5 \times 10^3~\mathrm{g\,cm^{-3}\,s^{-2}}$,  $\Theta_h \simeq 16.86$, 
$n_h \simeq 10^6~\mathrm{cm^{-3}}$, $T_h \simeq 10^{11}~\mathrm{K}$ which is applicable to the plasma  surrounding the $\text{M87}^*$~\cite{EventHorizonTelescope:2019dse}. In the following analysis, we fix $\epsilon=20$~\cite{Zhang:2024lsf}, corresponding to a regime where proton thermal motion remains nonrelativistic.

Considering the ideal MHD condition~\cite{Ruffini:1975ne} and the stationary, axisymmetric nature of the accretion flow, the gauge potential of the magnetic field is taken to be independent of both time and azimuthal coordinates. Furthermore, applying Maxwell’s equations yields~\cite{Ruffini:1975ne,Hou:2023bep}:
\begin{equation}
\label{beq}
B^\mu=\frac{\partial_\theta A_\phi}{\sqrt{-g} u^r}\left(\left(u_t+\Omega_B u_\phi\right) u^\mu+\delta_t^\mu+\Omega_B \delta_\phi^\mu\right)\,,
\end{equation}
where $\Omega_B$ In this expression, $\Omega_B$ denotes the angular velocity of the magnetic field lines. Since $B^i$ is parallel to $u^i$, the field is frozen into the fluid streamlines. In this work, we adopt the simplification $\Omega_B = 0$.  
The function
$\partial_\theta A_\phi$ can be chosen to adopt a split monopole configuration~\cite{Blandford:1977ds}:
\begin{equation}
\partial_\theta A_\phi=\Phi_0 \operatorname{sign}(\cos \theta) \sin \theta\,.
\end{equation}

Finally, based on the methodology introduced above, we present the black hole images illuminated by the BAAF disk with anisotropic synchrotron emission. In this model, the anisotropy of the emission naturally gives rise to polarized radiation. We analyze the resulting intensity and polarization distributions, with particular emphasis on how the black hole’s spin and frame-dragging effects manifest themselves in the observed polarization patterns.

\subsection{Anisotropic Radiation}
\begin{figure*}[htbp]
  \centering
  \begin{subfigure}{0.23\textwidth}
    \includegraphics[width=3.5cm,height=3cm]{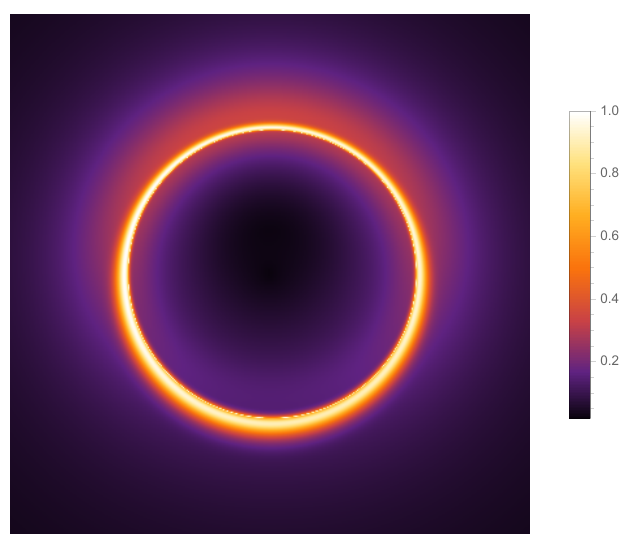}
    \caption{$a=0.1,\alpha=0.2$}
  \end{subfigure}
  \begin{subfigure}{0.23\textwidth}
    \includegraphics[width=3.5cm,height=3cm]{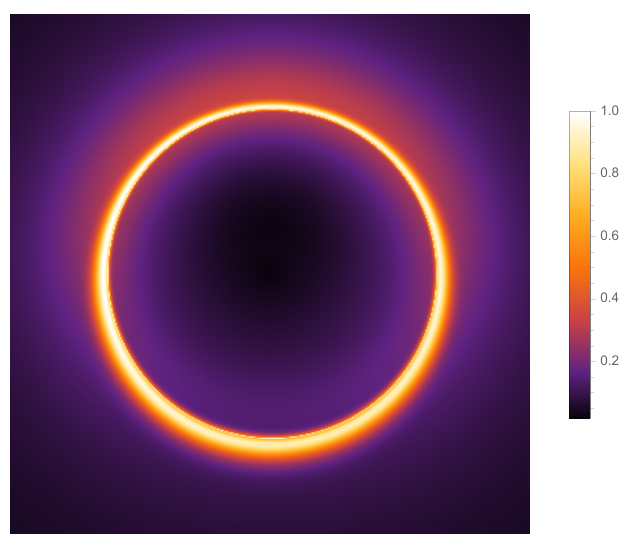}
   \caption{$a=0.1,\alpha=0.4$}
  \end{subfigure}
  \begin{subfigure}{0.23\textwidth}
    \includegraphics[width=3.5cm,height=3cm]{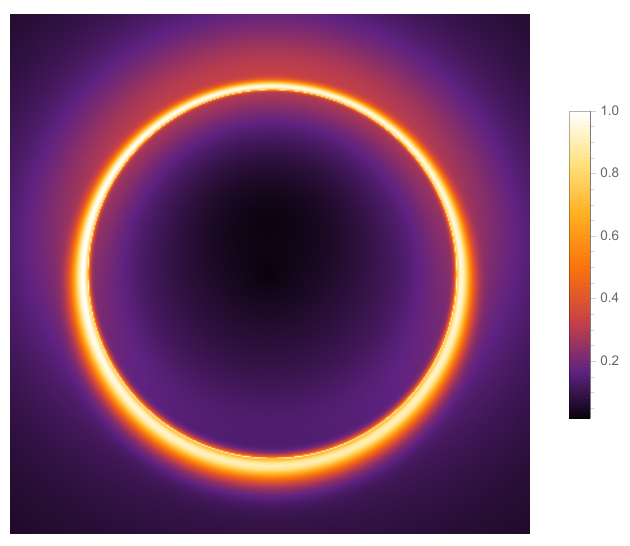}
  \caption{$a=0.1,\alpha=0.6$}
  \end{subfigure}
  \begin{subfigure}{0.23\textwidth}
    \includegraphics[width=3.5cm,height=3cm]{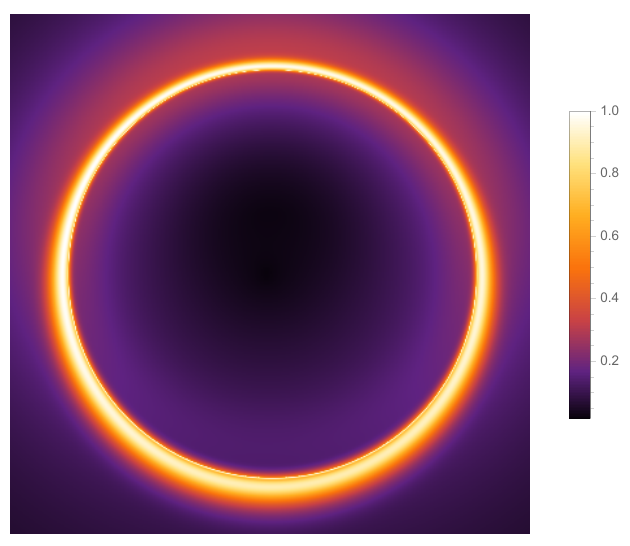}
    \caption{$a=0.1,\alpha=0.8$}
  \end{subfigure}
 \begin{subfigure}{0.23\textwidth}
    \includegraphics[width=3.5cm,height=3cm]{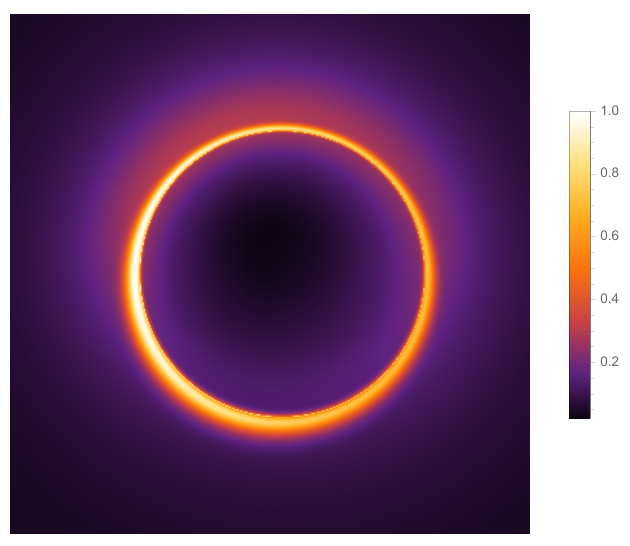}
    \caption{$a=0.5,\alpha=0.2$}
  \end{subfigure}
  \begin{subfigure}{0.23\textwidth}
    \includegraphics[width=3.5cm,height=3cm]{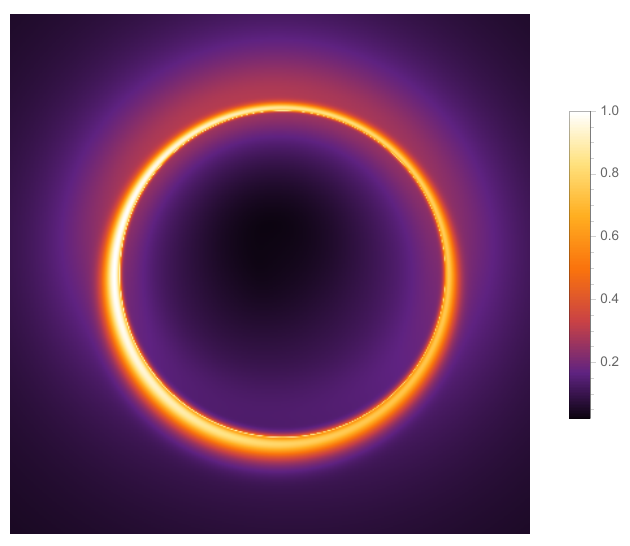}
    \caption{$a=0.5,\alpha=0.4$}
  \end{subfigure}
   \begin{subfigure}{0.23\textwidth}
    \includegraphics[width=3.5cm,height=3cm]{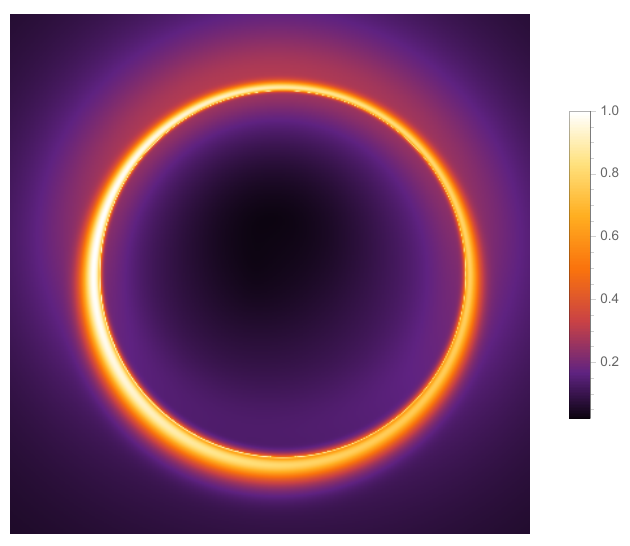}
    \caption{$a=0.5,\alpha=0.6$}
  \end{subfigure}
  \begin{subfigure}{0.23\textwidth}
    \includegraphics[width=3.5cm,height=3cm]{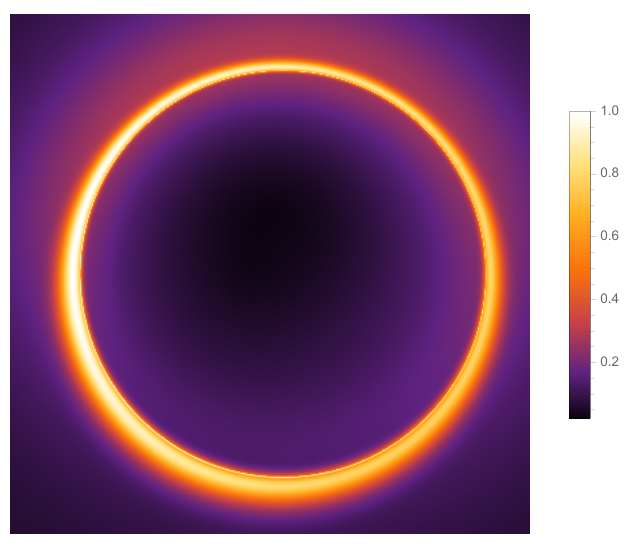}
    \caption{$a=0.5,\alpha=0.8$}
  \end{subfigure}
  \begin{subfigure}{0.23\textwidth}
    \includegraphics[width=3.5cm,height=3cm]{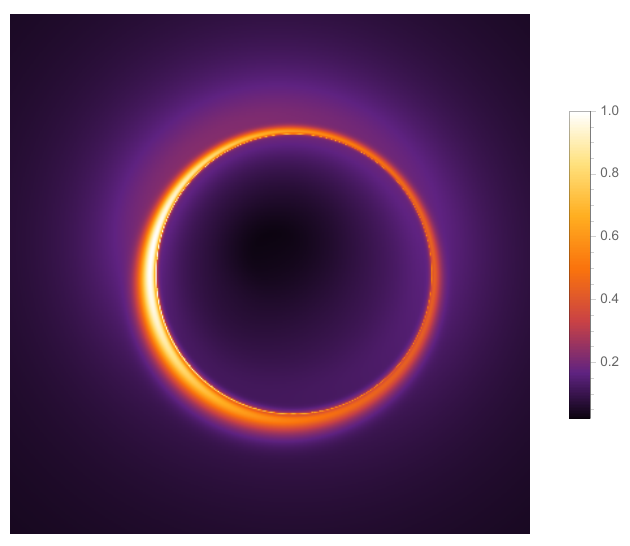}
    \caption{$a=0.9,\alpha=0.2$}
  \end{subfigure}
  \begin{subfigure}{0.23\textwidth}
    \includegraphics[width=3.5cm,height=3cm]{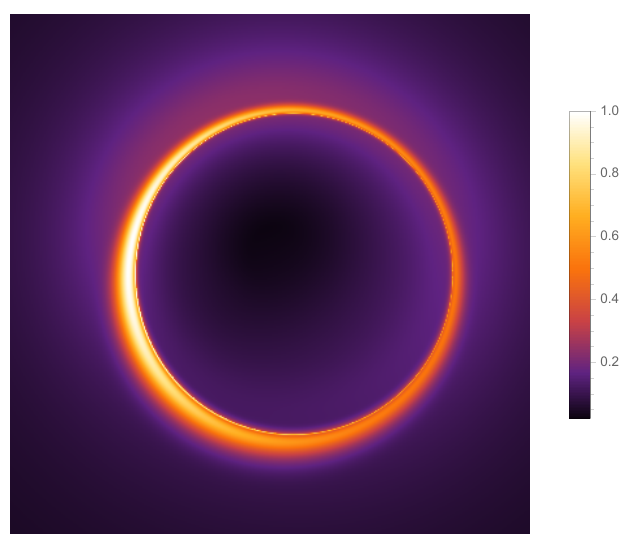}
    \caption{$a=0.9,\alpha=0.4$}
  \end{subfigure}
  \begin{subfigure}{0.23\textwidth}
    \includegraphics[width=3.5cm,height=3cm]{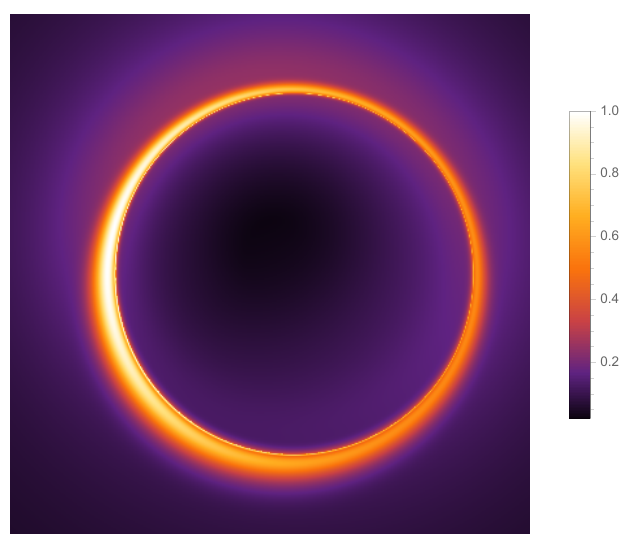}
    \caption{$a=0.9,\alpha=0.6$}
  \end{subfigure}
  \begin{subfigure}{0.23\textwidth}
    \includegraphics[width=3.5cm,height=3cm]{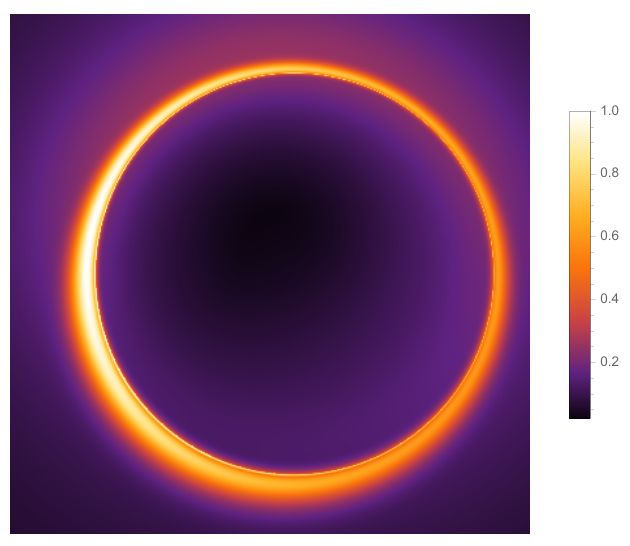}
    \caption{$a=0.9,\alpha=0.8$}
  \end{subfigure}
   \caption{Intensity maps of the Kerr–MOG black hole in the BAAF model with anisotropic synchrotron emission.
The accretion flow follows an infalling motion, and the observation frequency is fixed at $345~\mathrm{GHz}$.
From left to right, the spin parameter takes values $a=0.1,~0.5,~0.9$, while from top to bottom, the MOG parameter is set to $\alpha=0.2,~0.4,~0.6,~0.8$.
The inclination is $30^\circ$.}
   \label{fig:8}
\end{figure*}
\begin{figure*}[htbp]
  \centering
  \begin{subfigure}{0.45\textwidth}
    \includegraphics[width=6.5cm,height=3.5cm]{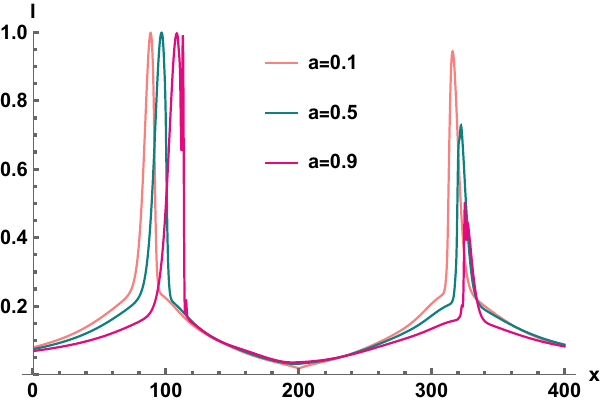}
    \caption{Horizontal}
  \end{subfigure}
   \begin{subfigure}{0.45\textwidth}
    \includegraphics[width=6.5cm,height=3.5cm]{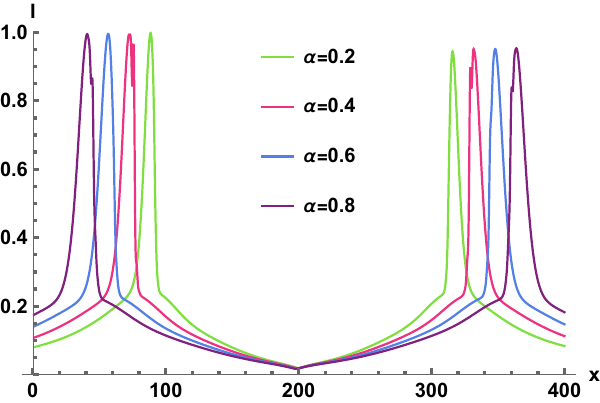}
    \caption{Horizontal}
  \end{subfigure}
 \\
  \begin{subfigure}{0.45\textwidth}
    \includegraphics[width=6.5cm,height=3.5cm]{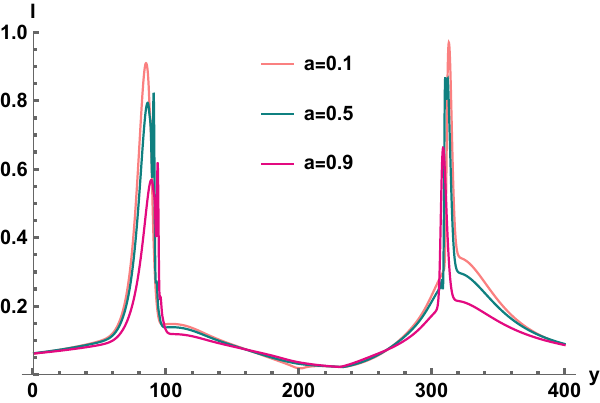}
    \caption{Vertical}
  \end{subfigure}
  \begin{subfigure}{0.45\textwidth}
    \includegraphics[width=6.5cm,height=3.5cm]{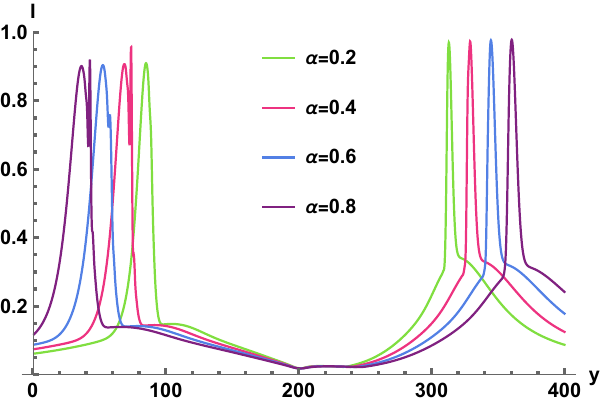}
    \caption{Vetical}
  \end{subfigure}
   \caption{Intensity distributions for the Kerr-MOG black hole in the BAAF model with anisotropic emission. 
     The accretion flow follows the infalling motion, and observer inclination is $30^\circ$.
    The left column shows profiles for varying spin $a$ with fixed $\alpha=0.2$, 
    while the right column shows profiles for varying MOG parameter $\alpha$ with fixed $a=0.1$.
    Upper and lower panels correspond to horizontal ($X$-axis) and vertical ($Y$-axis) cuts, respectively.}
   \label{fig:9}
\end{figure*}
\begin{figure*}[htbp]
  \centering
  \begin{subfigure}{0.23\textwidth}
    \includegraphics[width=3.5cm,height=3cm]{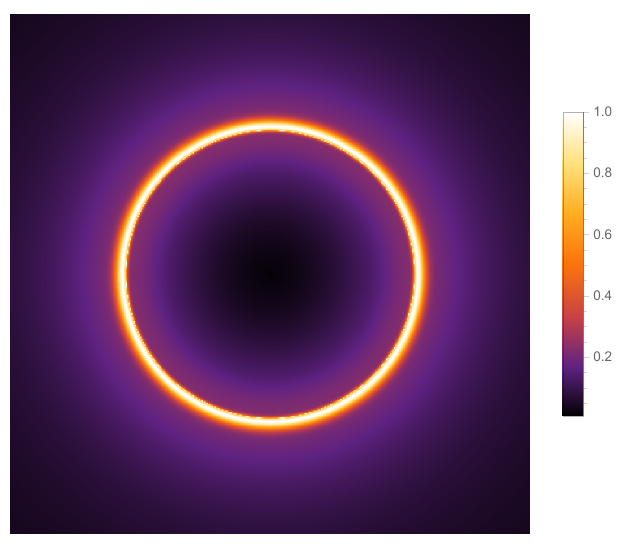}
    \caption{$\theta_o=1^\circ$}
  \end{subfigure}
  \begin{subfigure}{0.23\textwidth}
    \includegraphics[width=3.5cm,height=3cm]{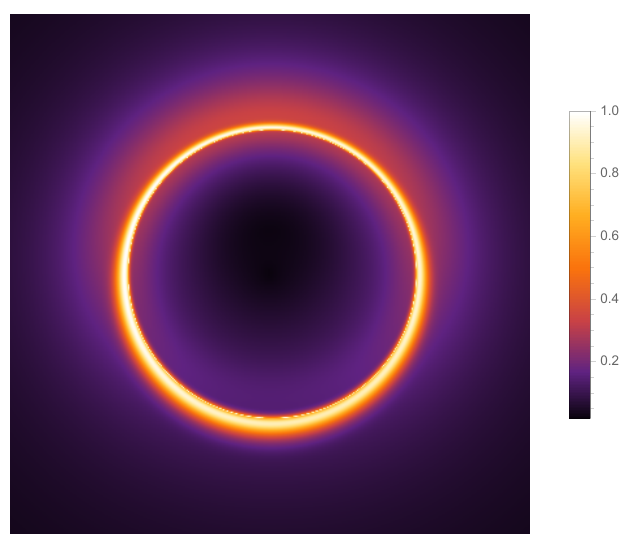}
   \caption{$\theta_o=30^\circ$}
  \end{subfigure}
  \begin{subfigure}{0.23\textwidth}
    \includegraphics[width=3.5cm,height=3cm]{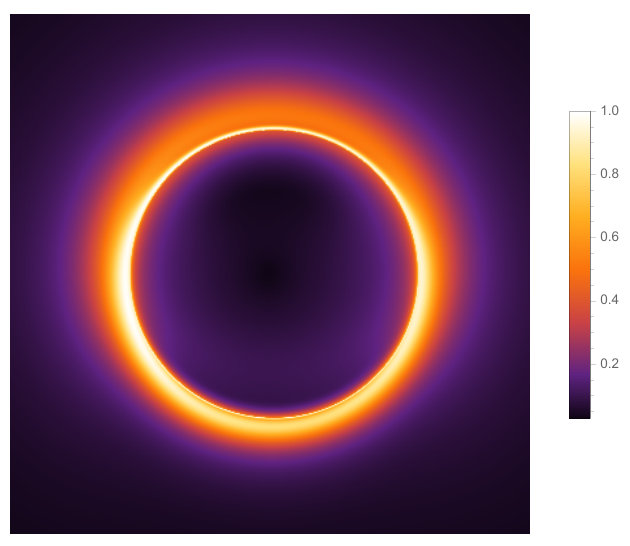}
   \caption{$\theta_o=60^\circ$}
  \end{subfigure}
  \begin{subfigure}{0.23\textwidth}
    \includegraphics[width=3.5cm,height=3cm]{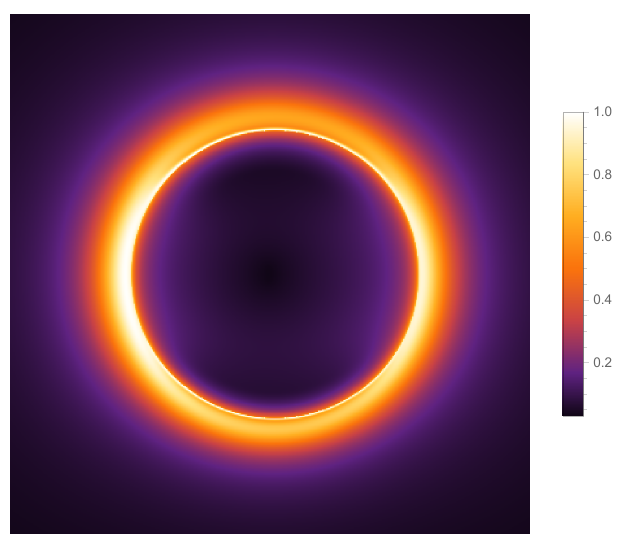}
    \caption{$\theta_o=80^\circ$}
  \end{subfigure}
   \caption{ Intensity maps in the BAAF model with anisotropic emission and infalling motion.
    Images are shown for inclinations $\theta_o = 1^\circ,\,30^\circ,\,60^\circ$, and $80^\circ$ (from left to right).
    The observation frequency is fixed at $345~\mathrm{GHz}$, and 
    the model parameters are $a=0.1$ and $\alpha=0.2$.}
   \label{fig:10}
\end{figure*}
Figure~\ref{fig:8} shows total intensity maps of the Kerr–MOG black hole computed with the BAAF disk model for different values of the spin $a$ and MOG parameter $\alpha$. The accretion flow is purely infalling, and the observation frequency is fixed at $345\,\mathrm{GHz}$. The associated horizontal and vertical intensity profiles are shown in Fig.~\ref{fig:9}, providing a quantitative comparison of the brightness distributions. Figure~\ref{fig:10} further illustrates the variation of the intensity morphology with the inclination angle.

Overall, the intensity distribution in the BAAF model exhibits trends with $a$, $\alpha$, and $\theta_o$ similar to those in the RIAF model. Compared with the RIAF case shown in Figs.~\ref{fig:5} and~\ref{fig:6}, the bright ring in the BAAF model is narrower, likely due to the disk being more optically thick for the chosen parameters, which enhances its separation from the primary image. Moreover, at large inclination angles, comparing panels (c) and (d) in Figs.~\ref{fig:10} and~\ref{fig:7}, the two distinct dark regions observed in the RIAF case do not appear, as parts of the BAAF disk are geometrically thinner in these regions for certain parameter choices.

\subsection{Polarization Structure}

In this subsection, we focus on the polarization signatures of the Kerr-MOG black hole surrounded by a BAAF thick disk, aiming to explore how the MOG parameter, flow dynamics and inclination angle influence the near-horizon polarization structure.
To facilitate a clear characterization of the correlation between the EVPA and the image's radial direction, we introduce the ``net EVPA'', $\chi_\text{net}$~\cite{Gelles:2021kti,Chen:2024jkm}, defined as
\begin{equation}
    \chi_\text{net}=\chi-\varphi\,,
\end{equation}
to denote the angle between the polarization  vector and the line from the origin of the screen to the image point ($\rho,\varphi$). This quantity encodes the pattern of the polarization. For instance, $\chi_\mathrm{net}=0^\circ$ or $180^\circ$ corresponds to a radial pattern in which the polarization vectors align with the radial direction $\hat{\rho}$, whereas $\chi_\mathrm{net}=90^\circ$ indicates a toroidal configuration with vectors oriented along the azimuthal direction $\hat{\varphi}$.

For polarization images with approximately central symmetry, we introduce the $\angle\beta_2$ to quantitatively describe the near-horizon linear polarization structure. The quantity $\angle\beta_2$ corresponds to the second azimuthal Fourier mode of the linear polarization map. The Fourier decomposition of the polarization vectors and the physical interpretation of $\angle\beta_2$ are discussed in detail in Appendix~\ref{appendix:A}.

\begin{figure*}[htbp]
  \centering
  \begin{subfigure}{0.48\textwidth}
    \includegraphics[width=7.8cm,height=6.5cm]{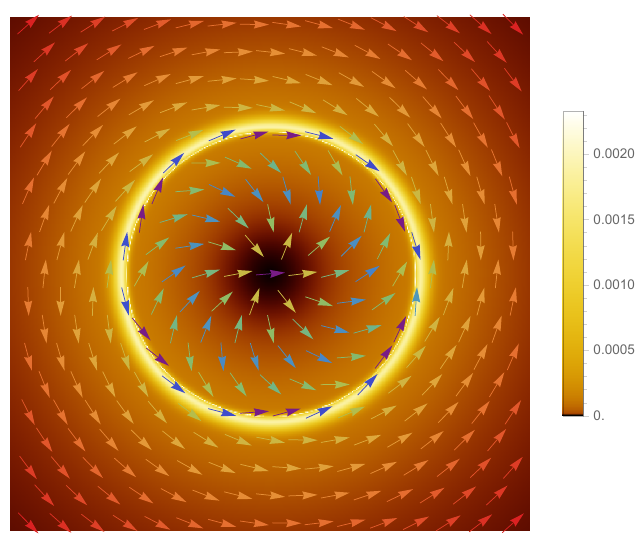}
    \caption{Stokes $I_o$}
  \end{subfigure}
  \begin{subfigure}{0.48\textwidth}
   \includegraphics[width=7.8cm,height=6.5cm]{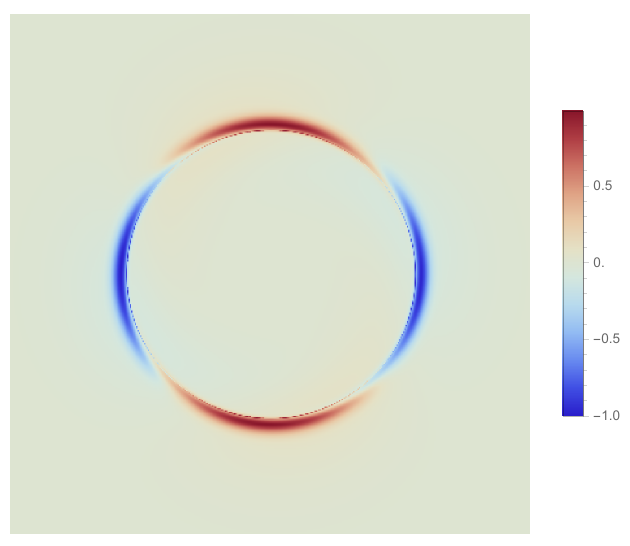}
    \caption{Stokes $Q_o$}
  \end{subfigure}\\
  \begin{subfigure}{0.48\textwidth}
     \includegraphics[width=7.8cm,height=6.5cm]{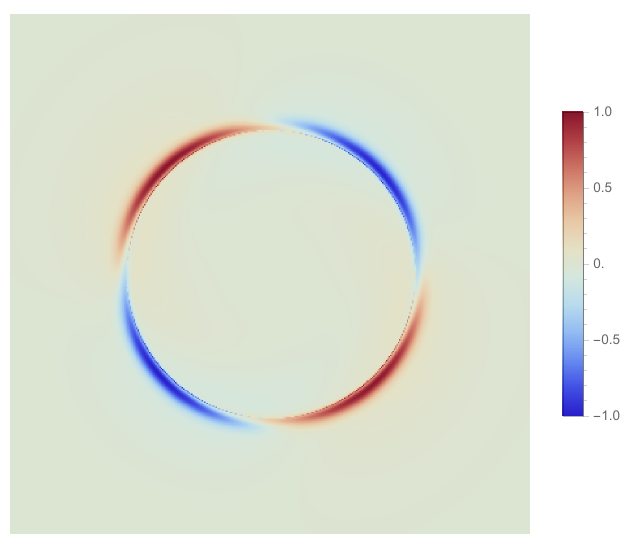}
    \caption{Stokes $U_o$}
  \end{subfigure}
   \begin{subfigure}{0.48\textwidth}
     \includegraphics[width=7.8cm,height=6.5cm]{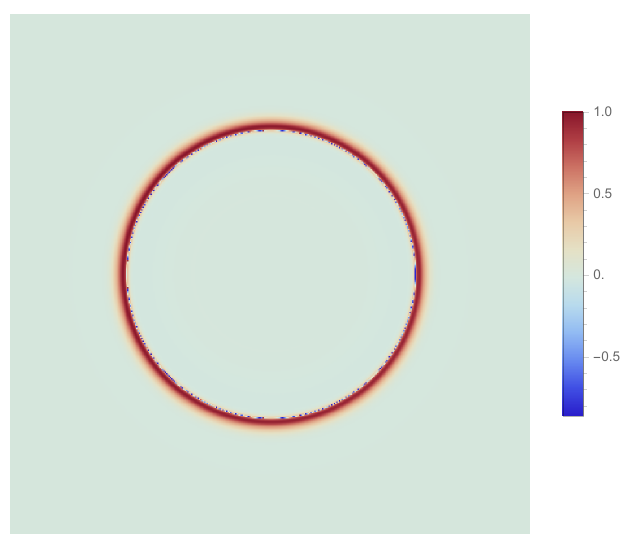}
    \caption{Stokes $V_o$}
  \end{subfigure}
   \caption{The resulting Stokes parameters $I_o$, $Q_o$, $U_o$, $V_o$ under the BAAF disk model. The dynamics of the accretion flow is infalling motion, with fixed parameters $a=0.9\,,\alpha=0.2\,, \theta_o= 30^\circ$.}
   \label{fig:11}
\end{figure*}
Figure~\ref{fig:11} presents a representative example of the observed Stokes parameters 
$I_o$, $Q_o$, $U_o$, and $V_o$. The arrows in panel (a) indicating the EVPA, $\chi$, and colors representing the linear polarization degree, $P_o$. The accretion flow is assumed to be purely infalling, with the parameters fixed at 
$a = 0.9$, $\alpha = 0.2$, and an observer inclination angle of $\theta_o = 30^\circ$. Since the EVPA is predominantly perpendicular to the magnetic field $B^\mu$, 
the polarization pattern provides insight into the field geometry. 
At large distances from the black hole, the inferred magnetic field is approximately radial, 
while closer to the horizon, frame-dragging effects become dominant, 
progressively twisting the magnetic field into a more azimuthal configuration. 
Due to the magnetic flux freezing condition in ideal MHD, the magnetic field remains largely aligned with the fluid motion; hence, the observed EVPA rotation also reflects the black hole’s frame dragging on the accreting plasma. 

\begin{figure*}[htbp]
  \centering
  \begin{subfigure}{0.23\textwidth}
    \includegraphics[width=3.7cm,height=3.5cm]{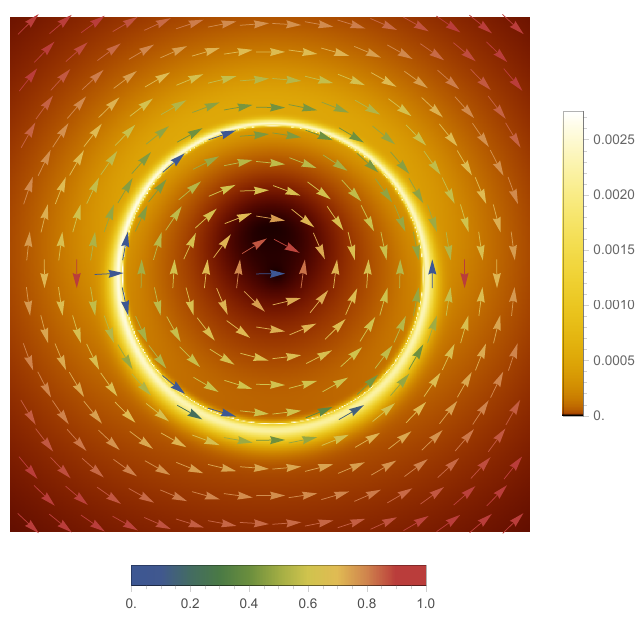}
    \caption{$a=0.1,\alpha=0.2$}
  \end{subfigure}
  \begin{subfigure}{0.23\textwidth}
     \includegraphics[width=3.7cm,height=3.5cm]{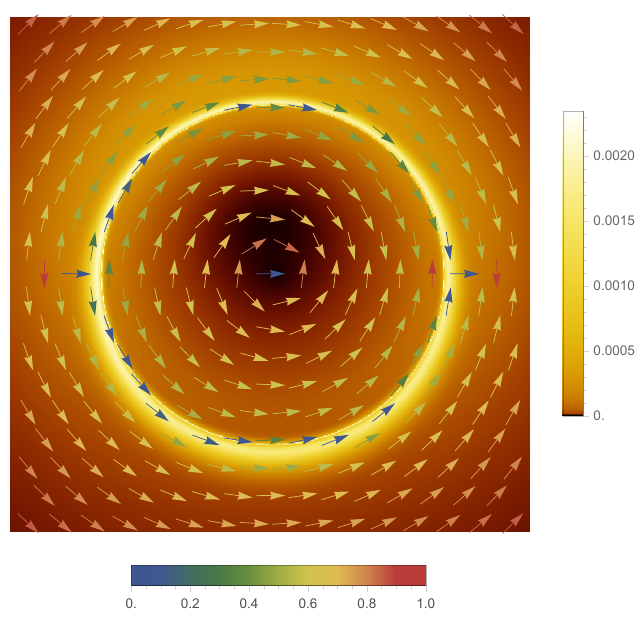}
   \caption{$a=0.1,\alpha=0.4$}
  \end{subfigure}
  \begin{subfigure}{0.23\textwidth}
     \includegraphics[width=3.7cm,height=3.5cm]{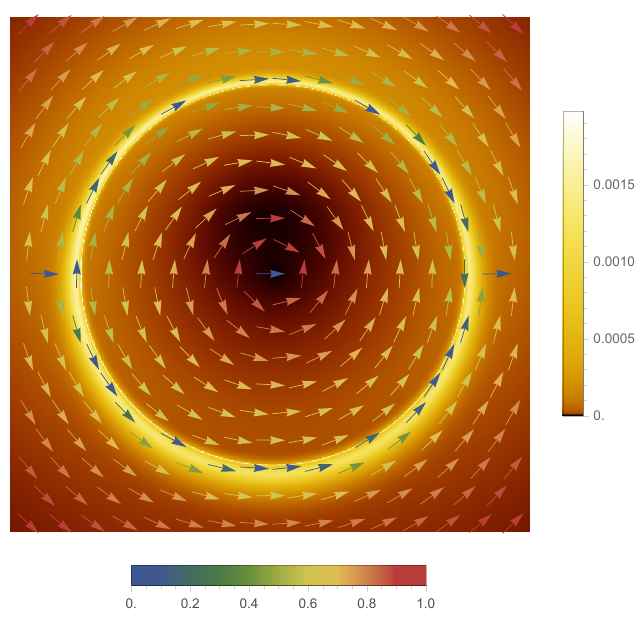}
  \caption{$a=0.1,\alpha=0.6$}
  \end{subfigure}
  \begin{subfigure}{0.23\textwidth}
     \includegraphics[width=3.7cm,height=3.5cm]{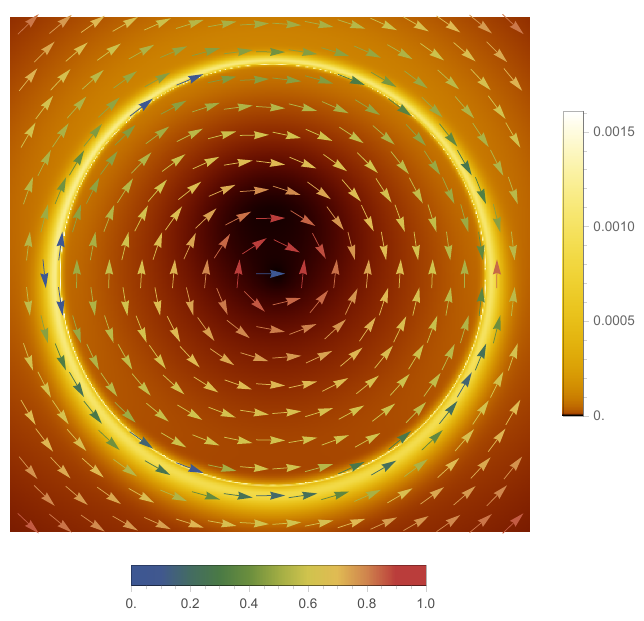}
    \caption{$a=0.1,\alpha=0.8$}
  \end{subfigure}
 \begin{subfigure}{0.23\textwidth}
      \includegraphics[width=3.7cm,height=3.5cm]{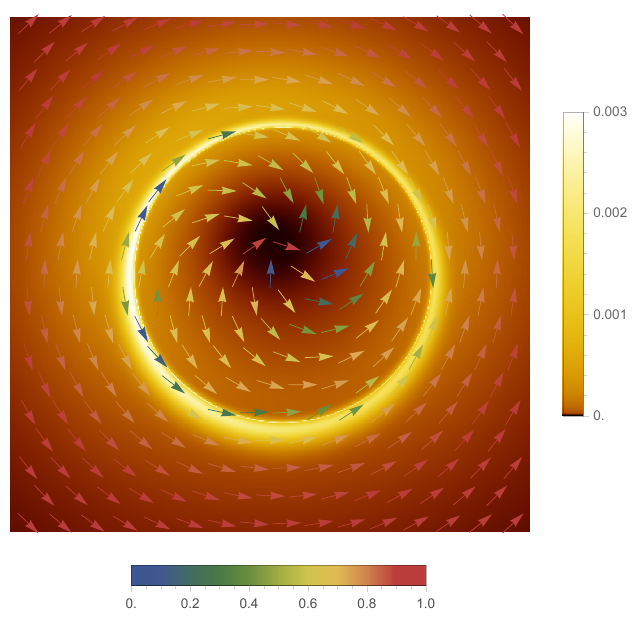}
    \caption{$a=0.5,\alpha=0.2$}
  \end{subfigure}
  \begin{subfigure}{0.23\textwidth}
      \includegraphics[width=3.7cm,height=3.5cm]{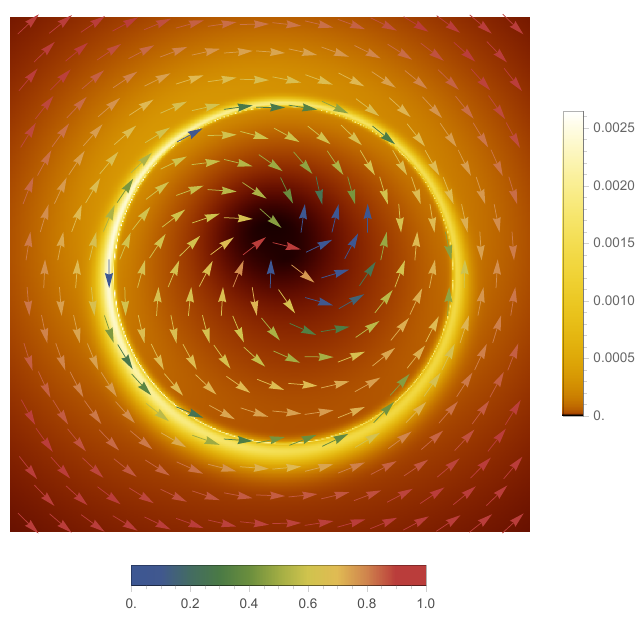}
    \caption{$a=0.5,\alpha=0.4$}
  \end{subfigure}
   \begin{subfigure}{0.23\textwidth}
      \includegraphics[width=3.7cm,height=3.5cm]{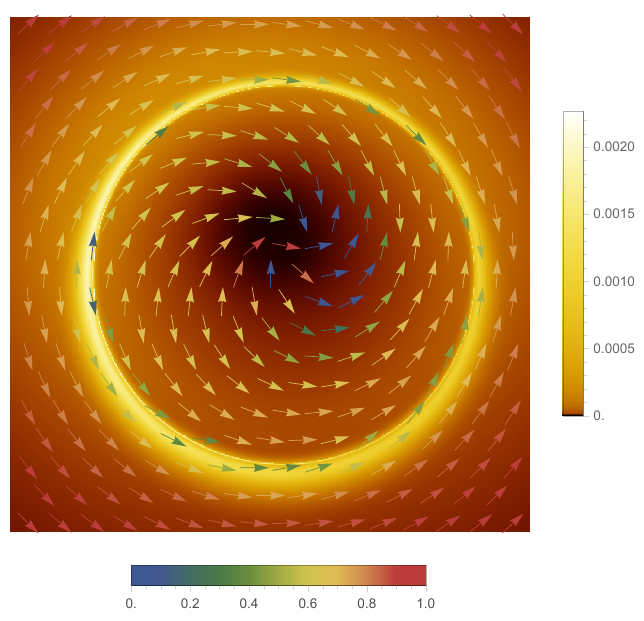}
    \caption{$a=0.5,\alpha=0.6$}
  \end{subfigure}
  \begin{subfigure}{0.23\textwidth}
     \includegraphics[width=3.7cm,height=3.5cm]{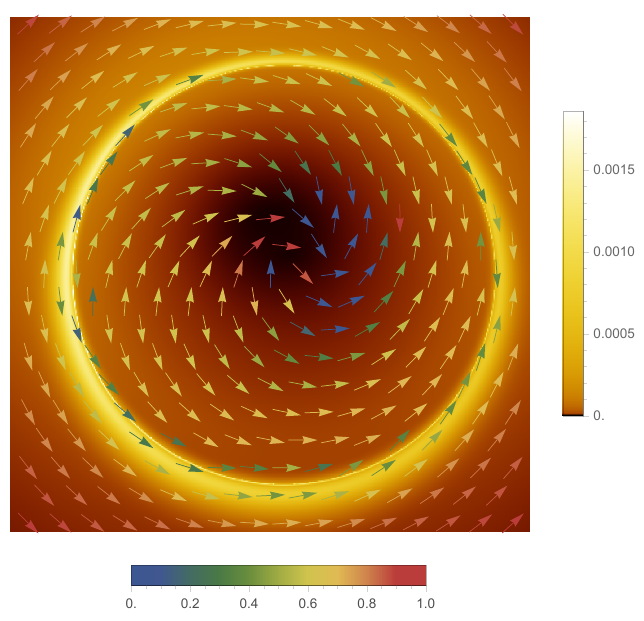}
    \caption{$a=0.5,\alpha=0.8$}
  \end{subfigure}
  \begin{subfigure}{0.23\textwidth}
    \includegraphics[width=3.7cm,height=3.5cm]{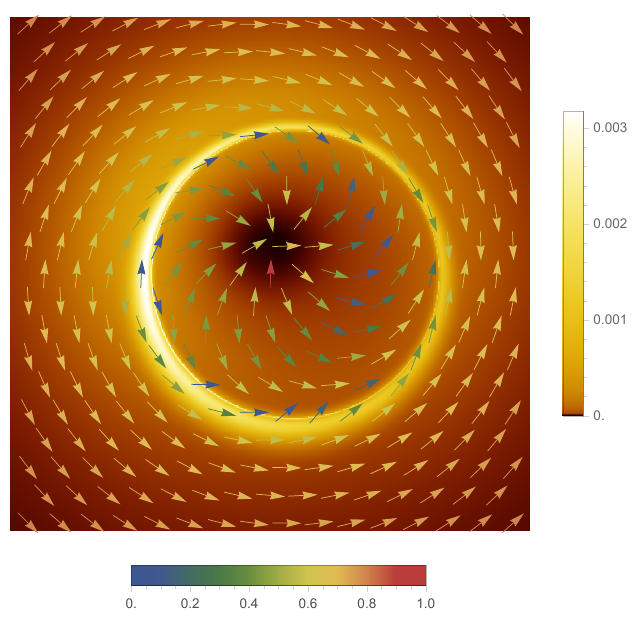}
    \caption{$a=0.9,\alpha=0.2$}
  \end{subfigure}
  \begin{subfigure}{0.23\textwidth}
    \includegraphics[width=3.7cm,height=3.5cm]{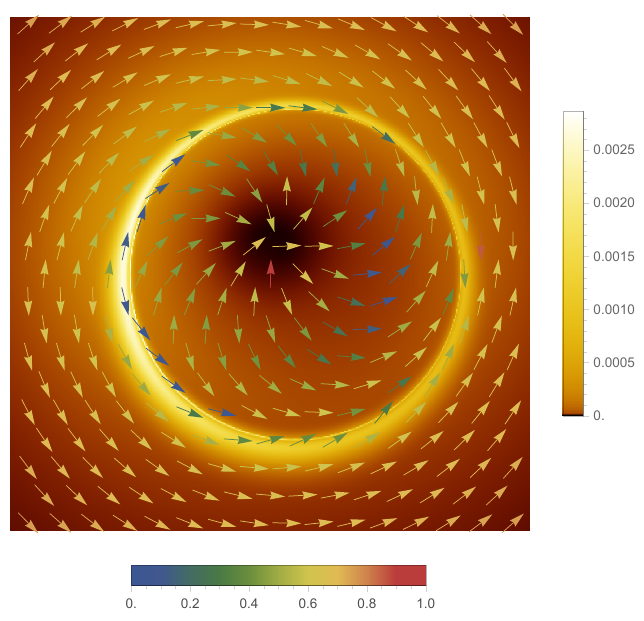}
    \caption{$a=0.9,\alpha=0.4$}
  \end{subfigure}
  \begin{subfigure}{0.23\textwidth}
     \includegraphics[width=3.7cm,height=3.5cm]{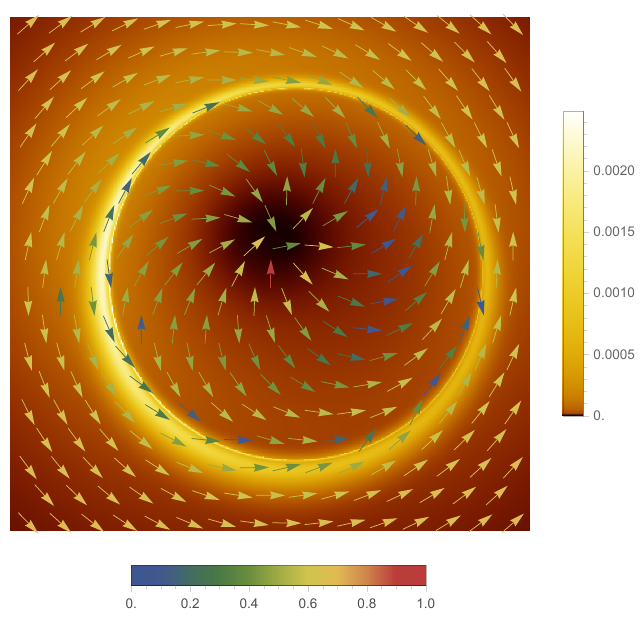}
    \caption{$a=0.9,\alpha=0.6$}
  \end{subfigure}
  \begin{subfigure}{0.23\textwidth}
     \includegraphics[width=3.7cm,height=3.5cm]{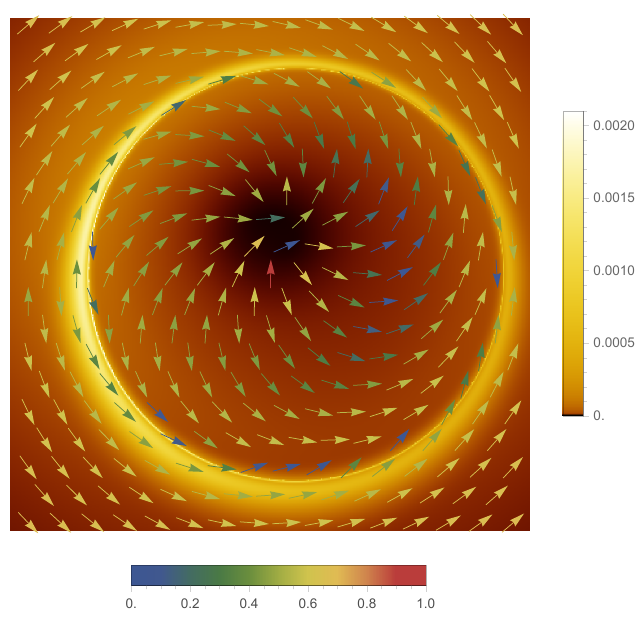}
    \caption{$a=0.9,\alpha=0.8$}
  \end{subfigure}
   \caption{ Polarized images of the Kerr-MOG black hole obtained from the BAAF disk model with anisotropic synchrotron emission.
The accretion flow follows an infalling motion, and the observer’s inclination angle is fixed at $30^\circ$. Each panel corresponds to a specific combination of the black hole spin $a$ (increasing from left to right: $0.1$, $0.5$, $0.9$) and the MOG parameter $\alpha$ (increasing from top to bottom: $0.2$, $0.4$, $0.6$, $0.8$).}
   \label{fig:12}
\end{figure*}
In Fig.~\ref{fig:12}, we present the polarization maps of the Kerr-MOG black hole 
for various combinations of the spin parameter $a$ and the MOG parameter $\alpha$. The accretion flow is assumed to be purely infalling, and the observer’s inclination angle is fixed at $\theta_o=30^\circ$. As the spin $a$ increases, the polarization vectors exhibit a stronger azimuthal twisting near the photon ring, indicating an enhanced frame-dragging effect. 
Meanwhile, by inspecting each row, we find that as $\alpha$ increases, the location where the polarization vectors begin to exhibit azimuthal twisting moves farther from the black hole. 
This behavior implies that a larger $\alpha$ effectively strengthens the frame-dragging effect in the Kerr-MOG spacetime.
Overall, the EVPA patterns transition from predominantly azimuthal (ring-like) orientations in the outer region to radial orientations close to the black hole, consistent with the magnetic flux freezing condition in ideal MHD.

\begin{figure*}[htbp]
  \centering
  \begin{subfigure}{0.23\textwidth}
    \includegraphics[width=3.7cm,height=3.5cm]{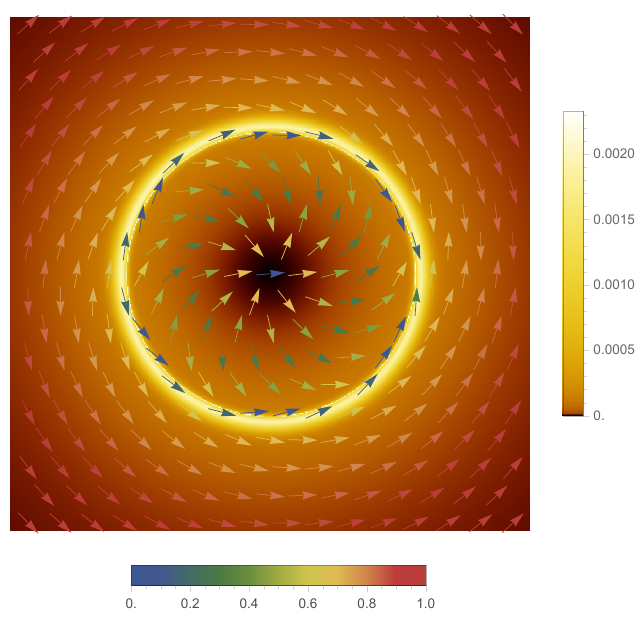}
    \caption{$\theta_o=1^\circ$}
  \end{subfigure}
  \begin{subfigure}{0.23\textwidth}
     \includegraphics[width=3.7cm,height=3.5cm]{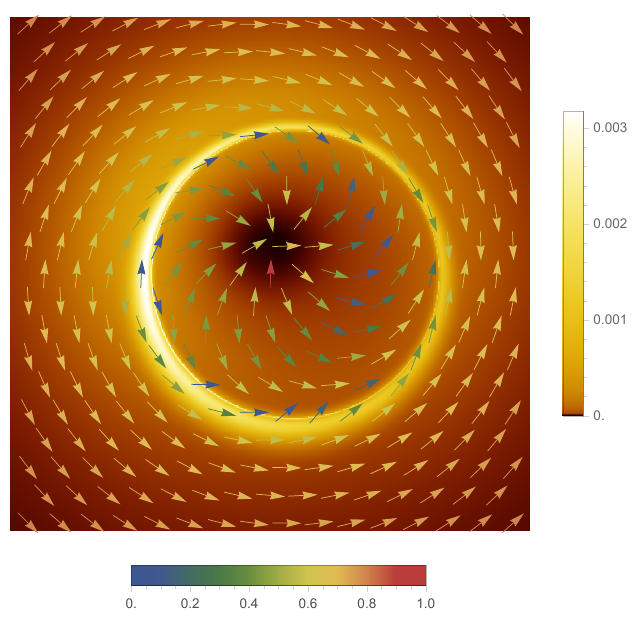}
   \caption{$\theta_o=30^\circ$}
  \end{subfigure}
  \begin{subfigure}{0.23\textwidth}
    \includegraphics[width=3.7cm,height=3.5cm]{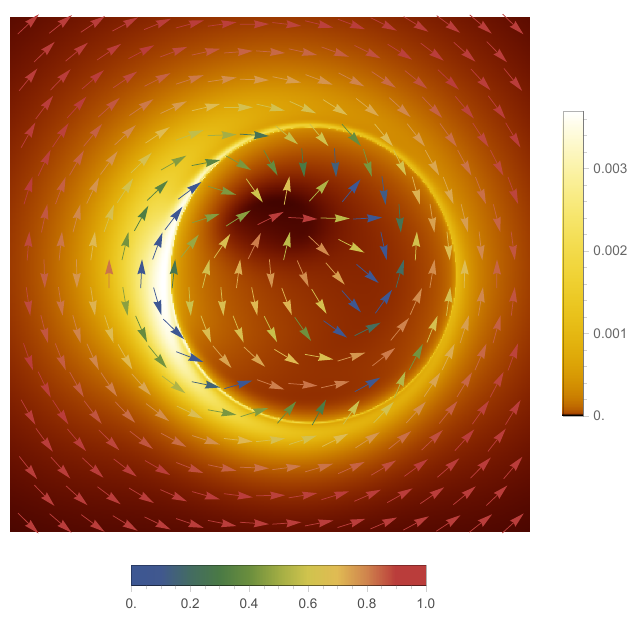}
   \caption{$\theta_o=60^\circ$}
  \end{subfigure}
  \begin{subfigure}{0.23\textwidth}
    \includegraphics[width=3.7cm,height=3.5cm]{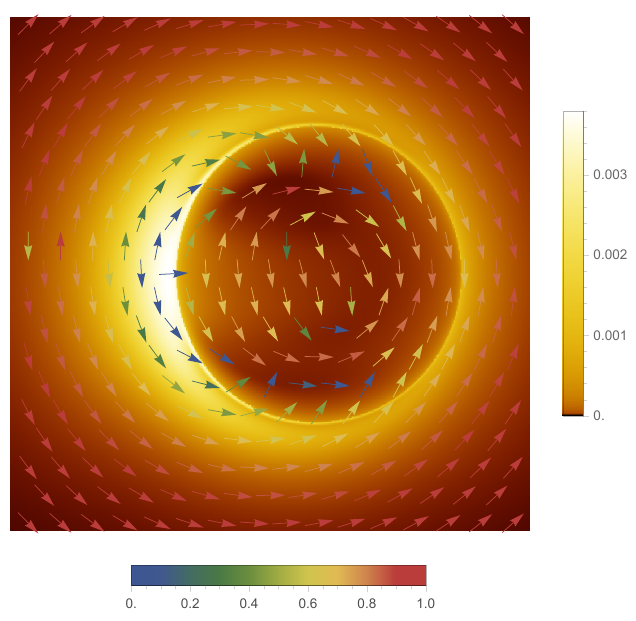}
    \caption{$\theta_o=80^\circ$}
  \end{subfigure}\\
  \begin{subfigure}{0.48\textwidth}
    \includegraphics[width=0.9\linewidth]{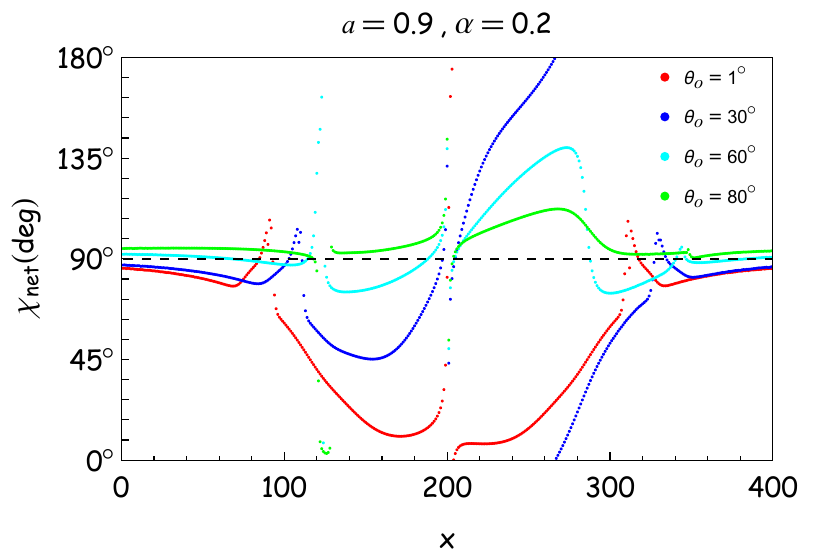}
    \caption{Horizontal}
  \end{subfigure}
  \begin{subfigure}{0.48\textwidth}
    \includegraphics[width=0.9\linewidth]{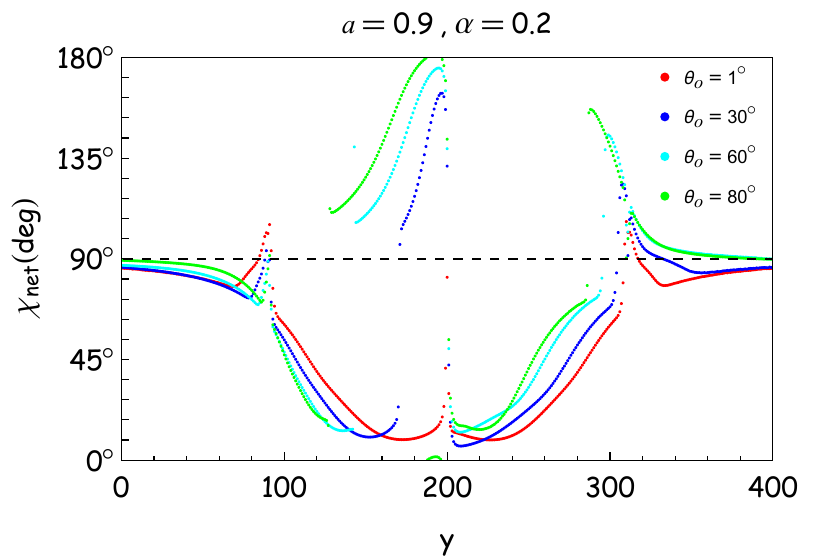}
    \caption{Vertical}
  \end{subfigure}
   \caption{Upper row: polarized images in the BAAF model with anisotropic emission and infalling motion.
    Images are shown for observing inclinations $\theta_o = 1^\circ,\,30^\circ,\,60^\circ$, and $80^\circ$ (from left to right).
    The model parameters are $a=0.9$ and $\alpha=0.2$. Lower row: the horizontal (left) and vertical (right) $\chi_\text{net}$ cuts for different inclinations.}
   \label{fig:13}
\end{figure*}
In Fig.~\ref{fig:13}, we present the polarized images for different observing inclinations.  
The spin and MOG parameters are fixed at $a = 0.9$ and $\alpha = 0.2$.  
As the inclination $\theta_o$ increases, the overall polarization morphology changes significantly.  
At small inclinations ($\theta_o = 1^\circ$), the polarized pattern appears nearly axisymmetric.  
With increasing inclination, the distribution of polarization vectors becomes increasingly asymmetric,  
indicating that both lensing and frame-dragging effects strongly influence the observed EVPA distribution.  In particular, at high inclinations ($\theta_o = 80^\circ$), the $Y$-direction profile exhibits two distinct dark zones separated by a central bright region.  
This feature arises from the partial obscuration of the event horizon silhouette by the accretion flow.

In the lower row of panels, we quantitatively examine the horizontal (left) and vertical (right) cuts of the net polarization angle, $\chi_{\text{net}}$, for different inclinations.  
As $\theta_o$ increases, the polarization profiles become progressively more asymmetric along both the $X$ and $Y$ directions. As shown in panel (f), where the image center is located around pixel $Y \simeq 200$, the net polarization angle $\chi_{\text{net}}$ varies from $180^\circ$ to $90^\circ$ above the center,  
and from $0^\circ$ to $90^\circ$ below it, exhibiting a clear mirror symmetry with respect to the disk midplane.

To further investigate how the polarization image reflects the frame-dragging effect, we consider two types of accretion flows with different specific energies, $E = 1.06$ and $E = 5$, both corresponding to retrograde motion ($L < 0$) as described in Eq.\eqref{conical}. For reference, an infalling flow with $E = 1$ is also included for comparison.

The accretion streamlines we consider plunge inward from large radii, where $u^\phi < 0$. Under the influence of the black hole’s frame-dragging effect, $u^\phi$ gradually increases and eventually becomes positive, indicating the existence of a turning point where $u^\phi = 0$. Under the magnetic flux freezing condition, and given that the polarization vector in thermal synchrotron emission is perpendicular to the photon momentum and the magnetic field \cite{1979rpa..book.....R}, the presence of such a turning point implies a polarity reversal of the magnetic-field-induced polarization along the streamline. At this critical position, the EVPA becomes perpendicular to the radial direction on the observer’s screen~\cite{Wang:2025btn}, which, as discussed in Appendix~\ref{appendix:A}, corresponds to a $\pm180^\circ$ flip in $\angle\beta_2$.
\begin{figure*}[htbp]
  \centering
  \begin{subfigure}{0.23\textwidth}
    \includegraphics[width=3.7cm,height=3.5cm]{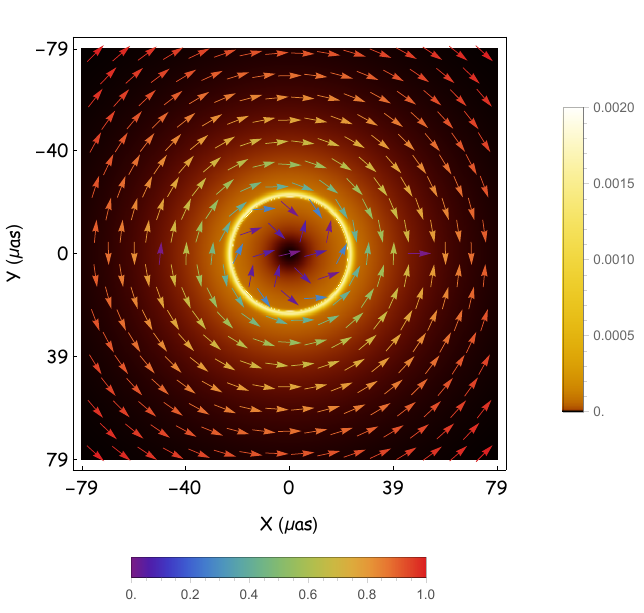}
    \caption{$\alpha=0.2\,,E=1.06$}
  \end{subfigure}
  \begin{subfigure}{0.23\textwidth}
   \includegraphics[width=3.7cm,height=3.5cm]{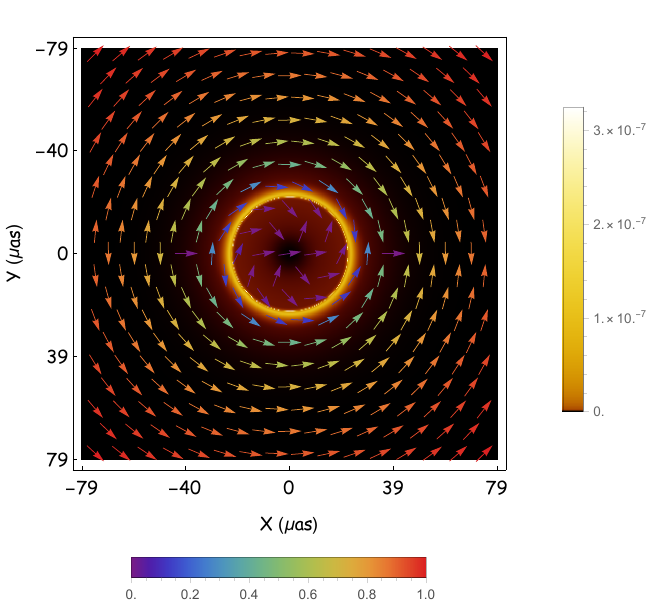}
    \caption{$\alpha=0.2\,,E=5$}
  \end{subfigure}
  \begin{subfigure}{0.23\textwidth}
     \includegraphics[width=3.7cm,height=3.5cm]{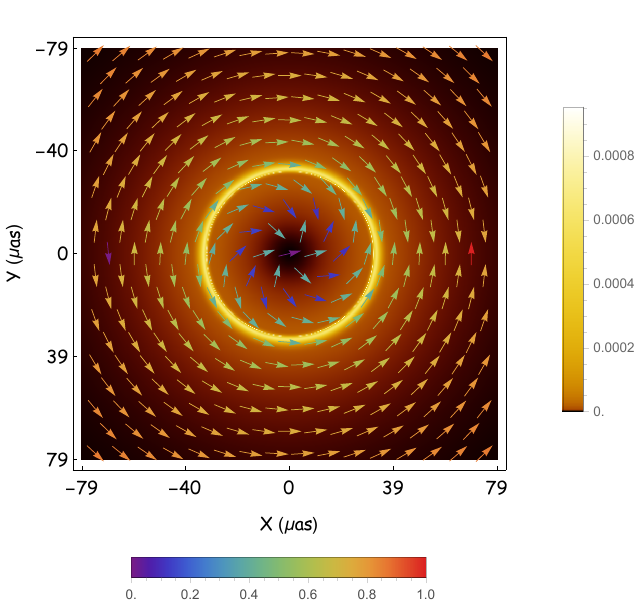}
    \caption{$\alpha=0.8\,,E=1.06$}
  \end{subfigure}
   \begin{subfigure}{0.23\textwidth}
     \includegraphics[width=3.7cm,height=3.5cm]{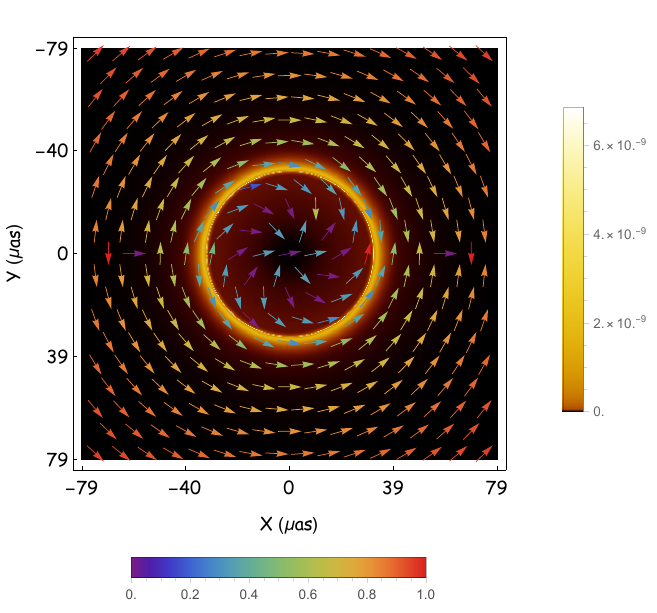}
    \caption{$\alpha=0.8\,,E=5$}
  \end{subfigure}\\
  \begin{subfigure}{0.48\textwidth}
     \includegraphics[width=0.9\linewidth]{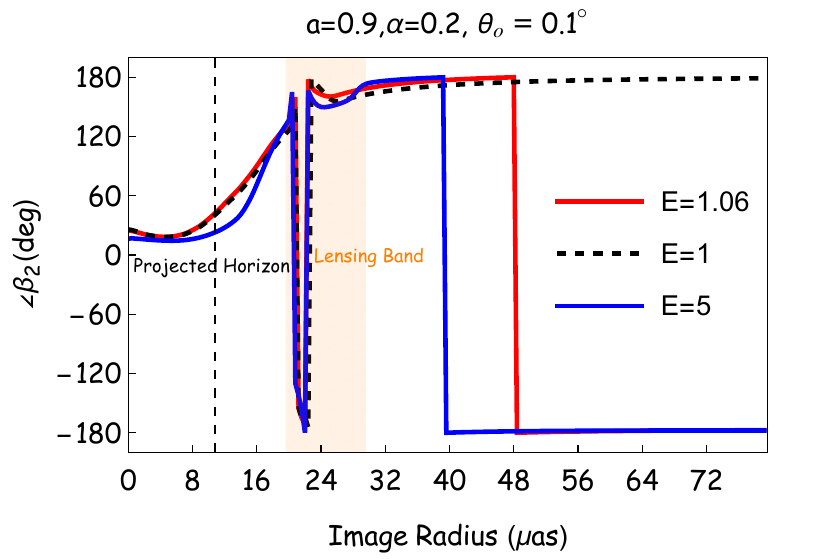}
    \caption{}
  \end{subfigure}
  \begin{subfigure}{0.48\textwidth}
     \includegraphics[width=0.9\linewidth]{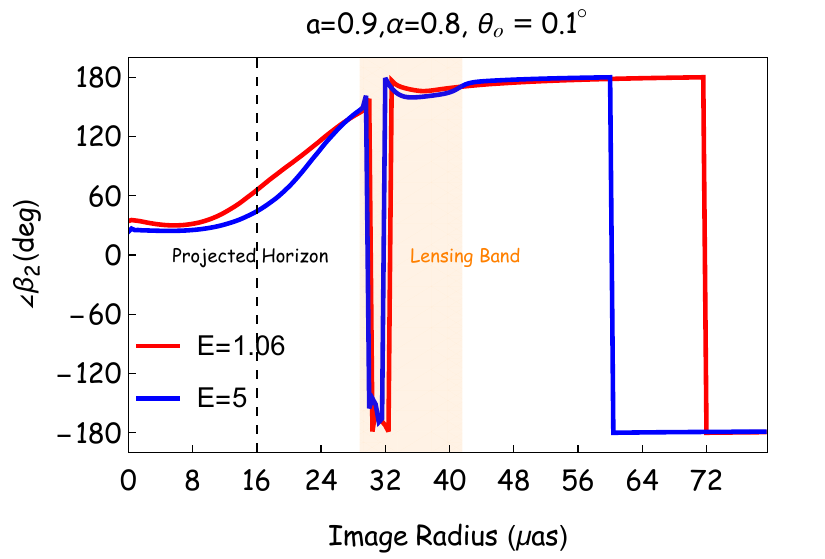}
    \caption{}
  \end{subfigure}
   \caption{Upper row: polarized images for different MOG parameters and flow energies. Lower row: $\angle\beta_2$ as a function of image radius for various accretion flow energies. The left panel corresponds to $\alpha = 0.2$, while the right panel shows results for $\alpha = 0.8$. The red, black dased, and blue curves represent $E=1.06$, $E=1$ (infalling case), and $E=5$, respectively. The vertical black dashed line denotes the ``projected horizon'' and the orange band indicates the ``lensing band''. All other parameters are fixed at $a=0.9\,,\theta_o=0.1^\circ$.}
   \label{fig:14}
\end{figure*}

We examine the EVPA distribution of these three accretion flows under different MOG parameters, viewed by a nearly polar observer. Figure\ref{fig:14} displays the polarized images for various combinations of $\alpha$ and flow energy $E$. To quantify the variation of polarization patterns, we compute the radial dependence of $\angle\beta_2$. Panels (e) and (f) correspond to $\alpha = 0.2$ and $\alpha = 0.8$, respectively. The red, black dashed, and blue curves denote accretion flows with specific energies $E = 1.06$, $E = 1$ (infalling case), and $E = 5$, respectively. The vertical black dashed line marks the ``projected horizon'', corresponding to the intersection of the event horizon with the equatorial plane. The orange shaded region denotes the ``lensing band'', representing the domain of higher-order lensed images.

As shown in Figs.~\ref{fig:14} (e)–(f), both retrograde ($L < 0$) accretion flows exhibit a distinct flip in $\angle\beta_2$, located outside the photon ring, whereas the infalling case does not show such a feature. Moreover, the higher the flow energy, the closer the turning point lies to the black hole. Near the projected horizon, $\angle\beta_2$ tends to converge. This convergence arises because, although near-horizon polarization is expected (the intrinsic polarization angle in the geometrically thin limit is independent of fluid motion \cite{Hou:2024qqo,Wong:2025zuh}), additional astrophysical effects, such as Faraday rotation and disk thickness, can significantly modify the near-horizon polarization in thick disk models. Comparing panels (e) and (f), we find that for larger values of $\alpha$, the $\angle\beta_2$ flip occurs at a larger image radius, implying that the frame-dragging effect becomes more pronounced as $\alpha$ increases. Furthermore, the distinct behavior of $\angle\beta_2$ near the projected horizon indicates that the MOG parameter $\alpha$ also has a noticeable impact on the polarization structure in the near-horizon region.

\section{Summary and discussion}\label{sec5}

In this work, we investigated the imaging and polarization properties of stationary, axisymmetric black holes in the Kerr–MOG spacetime surrounded by geometrically thick accretion flows. We first reviewed the essential features of the Kerr–MOG black hole, including its event horizon and geodesics. A class of fluid four-velocity configurations satisfying $u^{\theta}=0$, referred to as conical solutions, was then considered. For the retrograde case ($L<0$), we analytically derived the turning point where $u^{\phi}=0$, which arises due to the frame-dragging effect of the rotating black hole. The asymptotic behavior of this turning point for $E \to \infty$ was also obtained, showing dependence solely on the MOG parameter $\alpha$ but not on the spin $a$. As a special subclass of these conical solutions, we further examined the purely infalling motion ($E=1$). The synchrotron radiation of thermal electrons in the magnetofluid was then propagated to a distant observer to produce the black hole images. In practice, we numerically solved the null geodesic equations and the polarized radiative transfer equations to obtain the resulting images. Subsequently, we explored two representative models of geometrically thick accretion flows: a phenomenological RIAF model and an analytical BAAF model.

For the RIAF model, we analyzed both isotropic and anisotropic emission cases with purely infalling motion for various spacetime parameters $a$ and $\alpha$, under different observer inclinations and observation frequencies. The resulting images show that both the bright ring and the central dark region expand with increasing MOG parameter $\alpha$, while they shrink with increasing black hole spin $a$. As the inclination angle increases, the intensity distribution becomes noticeably asymmetric, and for observers near the equatorial plane, high-latitude emission partially fills the central darkness, producing distinct dark regions. The intensity distribution also varies with observation frequency: at higher frequencies, the bright ring becomes more pronounced due to the reduced optical depth and the dominance of emission from regions closer to the event horizon. When accounting for anisotropic emission, i.e., the dependence on the pitch angle $\theta_B$, the intensity maps show directional variations; in this work, for an observer near the equator, the bright ring is stretched along the vertical ($Y$) direction.

Finally, we analyzed the imaging and polarization characteristics of the BAAF disk.
For the intensity distribution, we again considered purely infalling fluids with anisotropic synchrotron emission, in order to facilitate a direct comparison with the RIAF model.
The intensity distributions broadly similar to those of the RIAF model with respect to $a$, $\alpha$, and $\theta_o$. The bright ring in the BAAF case is sharper, likely due to a higher optical thickness, and at large inclinations, the two distinct dark regions seen in the RIAF model are less pronounced because portions of the BAAF disk are geometrically thinner.

We then focused on the polarized images properties of the BAAF model. The polarized intensity closely traces the total brightness distribution, such that regions with larger total intensity also exhibit stronger polarized emission. The linear polarization degree is significantly suppressed in the bright ring. Moreover, both the polarization degree and the EVPA display a clear dependence on the black hole parameters $a$, $\alpha$, and the observer inclination $\theta_o$. To quantify the near-horizon polarization patterns, we introduced the ``net EVPA'', $\chi_{\text{net}} = \chi - \varphi$, and the second azimuthal Fourier mode, $\angle\beta_2$.
Our results show that the EVPA distribution transitions smoothly from a radial configuration in the outer region to a toroidal pattern near the black hole, consistent with the magnetic flux freezing condition in ideal MHD.

As the spin $a$ increases, the polarization vectors exhibit stronger azimuthal twisting near the photon ring, reflecting enhanced frame dragging.
For larger $\alpha$, this twisting occurs at greater radii, indicating that the MOG parameter effectively amplifies the frame-dragging behavior.
With increasing inclination, the EVPA distribution becomes more asymmetric, exhibiting clear mirror symmetry about the disk midplane.
The evolution of $\angle\beta_2$ further confirms these trends: retrograde accretion flows display a distinct $\pm180^{\circ}$ flip of $\angle\beta_2$ outside the photon ring, whereas the purely infalling case does not.
The location of this flip depends on both the flow energy and the spacetime parameters.
Moreover, near the projected horizon, $\angle\beta_2$ tends to converge, indicating the presence of an intrinsic polarization angle that, however, can be affected by astrophysical processes such as Faraday rotation and disk thickness.
This intrinsic polarization angle also encodes information about the MOG parameter $\alpha$, suggesting that the polarized image of thick disk may serve as a promising observational probe of modified gravity in strong field regimes.

We conclude this paper with several prospects for future exploration. 
In our analysis of near-horizon polarization, we have restricted attention to specific parameter choices of the BAAF model, such as disk thickness and temperature. Although intrinsic near-horizon polarization may exist in Kerr-MOG black holes, various astrophysical effects can obscure these theoretical signatures. Therefore, disentangling the gravitational contributions from the plasma and radiative effects remains an important and challenging direction for future studies.

\section*{Acknowledgments}
We are grateful to Minyong Guo, Zhenyu Zhang, Yehui Hou, Jiewei Huang for insightful discussions. This work is supported by the National Natural Science Foundation of China (Grants Nos. 12375043,
12575069,12275004 and 12205013).

\appendix
\section{Decomposition of Linear Polarization}
\label{appendix:A}
\begin{figure}[ht]
    \centering
    \includegraphics[width=6in]{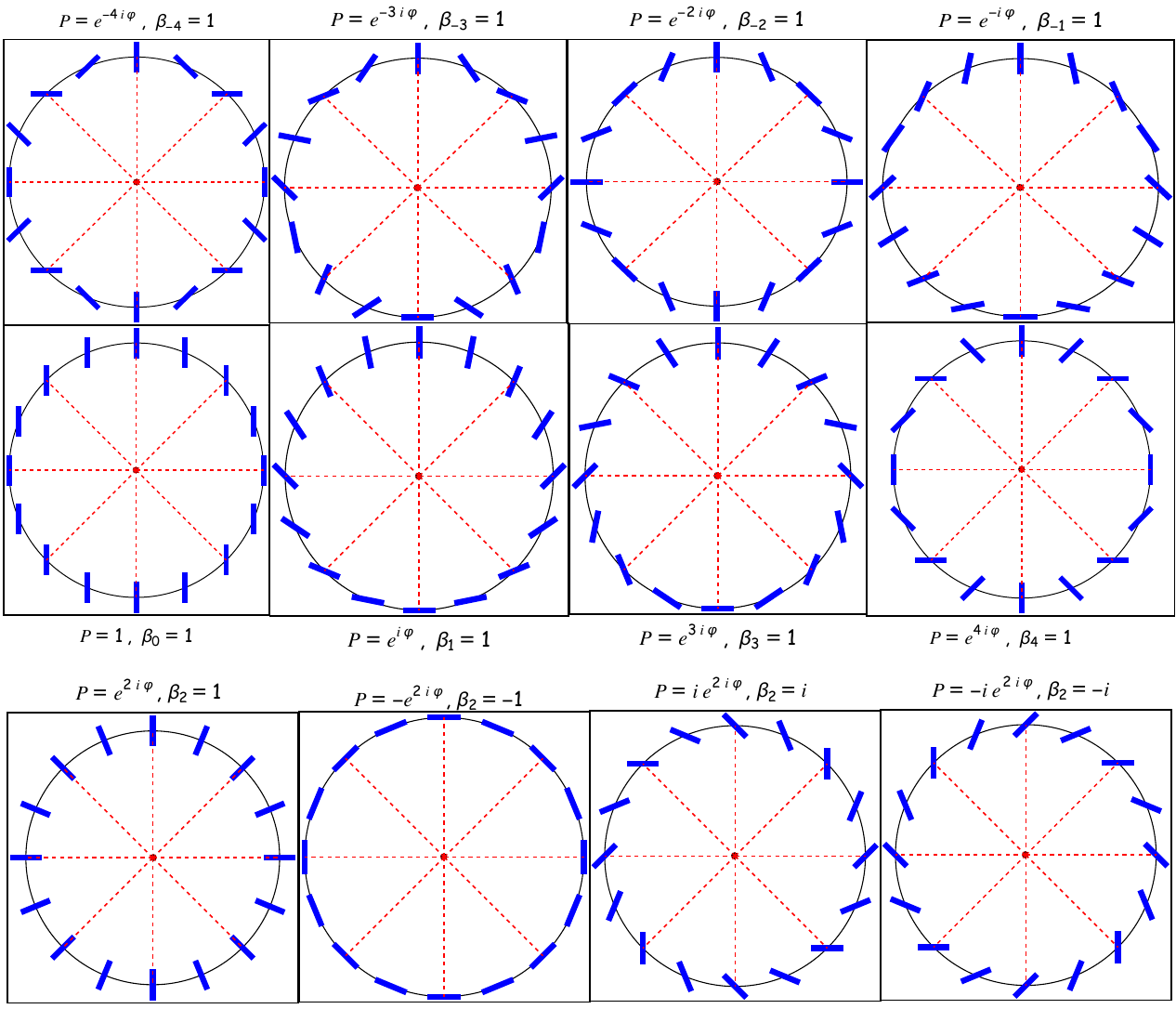}
    \caption{Examples of the EVPA for periodic polarization fields are plotted along a unit-radius ring, along with their corresponding \(\beta_m\). The first two rows present the results for \( -4 \leq m \leq 4 \) with \( \beta_m = 1 \,\), while the third row illustrates the cases for \( \beta_2 = \pm 1, \pm i \,\).}
    \label{fig:beta}
\end{figure}

Following \cite{Palumbo:2020flt}, we decompose the polarization field into a series of basis functions, given by  
\( P_m(\varphi) \equiv e^{i m \varphi} \). Within an annular region defined by the inner and outer radii  
\( \rho_{\min} \) and \( \rho_{\max} \), the Fourier coefficients corresponding to the integral Fourier modes are expressed as  
\bea  
\beta_m &=& \frac{1}{I_{\mathrm{ann}}} \int_{\rho_{\min}}^{\rho_{\max}} \int_0^{2\pi} P_o(\rho, \varphi) P_m^\ast(\varphi) \rho \, d\varphi \, d\rho \nn\\  
&=& \frac{1}{I_{\mathrm{ann}}} \int_{\rho_{\min}}^{\rho_{\max}} \int_0^{2\pi} P_o(\rho, \varphi) e^{-i m \varphi} \rho \, d\varphi \, d\rho \,,  
\eea  
with the normalization factor  
\bea  
I_{\mathrm{ann}} = \int_{\rho_{\min}}^{\rho_{\max}} \int_0^{2\pi} I(\rho, \varphi) \rho \, d\varphi \, d\rho \,,  
\eea  
where \( I(\rho, \varphi) \) denotes the total intensity brightness. In this work, we evaluate \( \beta_m \) for each pixel, meaning that we set the width of the ring to be the length of a pixel on the screen. The magnitude of \( \beta_m \) quantifies the strength of the corresponding mode, while its phase characterizes the average pointwise rotation of the polarization relative to a reference EVPA orientation. We define this reference as being vertically aligned along the \( \phi = 0 \) image axis, corresponding to the \( Y \)-axis direction.

Similar to Fig.\ 1 in \cite{Palumbo:2020flt}, we also present several illustrative examples of ring-valued linear polarization fields alongside their corresponding \( \beta_m \) in Fig.\ \ref{fig:beta}. The first two rows of the figure display the results for \( \beta_m = 1 \) with \( -4 \leq m \leq 4 \,\), which distinctly reveal the polarization distributions associated with these special modes, offering an intuitive understanding of how different values of \( \beta_m \) correspond to various polarization patterns.

Of particular interest is the \( m = 2 \) mode, as polarization patterns frequently exhibit rotational symmetry. In the third row of Fig.\ \ref{fig:beta}, we present the cases for \( \beta_2 = \pm 1, \pm i \). It is readily observed that when \( \beta_2 = 1 \), the polarization direction is aligned with the radial direction, whereas for \( \beta_2 = -1 \), it is perpendicular to the radial direction. Moreover, clockwise EVPA structures correspond to \( \operatorname{Im}(\beta_2) > 0 \,\), while counterclockwise patterns correspond to \( \operatorname{Im}(\beta_2) < 0 \,\).

By computing \( \beta_2 \), we obtain a qualitative understanding of the polarization distribution. In particular, when \( \beta_2 = -1 \), the phase satisfies \( \angle\beta_2 = 180^\circ \) or \( -180^\circ \), indicating that \( \beta_2 \) undergoes a phase flip.

\bibliographystyle{utphys}
\bibliography{reference}

\end{document}